\documentclass[a4paper, 11pt]{article}
\usepackage{jheppub} 
\usepackage{jheppub} 
\usepackage[utf8]{inputenc}
\usepackage{physics}
\usepackage{slashed}
\usepackage{cancel}
\usepackage{caption}
\usepackage{xcolor}
\usepackage{comment}
\usepackage{multirow}
\usepackage{graphics}
\usepackage{float}
\usepackage{cancel}
\usepackage{soul}
\usepackage{cases}
\usepackage{array}
\usepackage{mathtools}  
\usepackage{amsfonts}
\usepackage{hyperref}
\usepackage{amsmath}
\usepackage{amssymb}
\usepackage{tcolorbox}
\usepackage{tikz}
\usepackage{tikz-feynman}
\tikzfeynmanset{compat=1.1.0} 

\title{The $\mathcal{N}=1$ Super-Grassmannian for CFT$_3$ and a Foray on AdS and Cosmological Correlators}

\author{Aswini Bala,  Sachin Jain,  Dhruva K.S.,  Adithya A Rao}

\affiliation{Indian Institute of Science Education and Research, \\ Dr Homi Bhabha Road,  Pashan,  Pune,  India}

\emailAdd{aswini.bala@students.iiserpune.ac.in}
\emailAdd{sachin.jain@iiserpune.ac.in}
\emailAdd{k.s.dhruva@students.iiserpune.ac.in}
\emailAdd{adithya.arao@students.iiserpune.ac.in}

\abstract{We construct a Super-Grassmannian integral representation for $n-$point functions in  $\mathcal{N}=1$ SCFT$_3$. In this formalism,  conformal invariance,  supersymmetry,  and special superconformal invariance are implemented manifestly through (operator-valued) delta function constraints. An important feature of this framework is the fact that we obtain simple algebraic relations among component correlators, which enable us to determine any component correlator in terms of just one of the component correlators. In particular,  this formalism enables us to construct (A)dS$_4$ boundary correlators with contact diagrams from those that receive contributions purely from particle exchanges. We illustrate this by determining the (A)dS$_4$ Yang-Mills gluon four-point function from its gluino counterpart. Further,  we establish the flat-space limit in super-space,  finding a perfect agreement with existing flat-space results.}
\emailAdd{~}

\begin{document}

\maketitle

\section{Introduction}
Over the past decade,  the conformal bootstrap program has developed into a broad framework encompassing rigorous numerical bounds on the conformal field theory (CFT) data ~\cite{Poland:2022qrs},  analytic expansions for correlators in controlled kinematic limits~\cite{Fitzpatrick:2012yx, Komargodski:2012ek},  and progress 
in theories with additional symmetry~\cite{Maldacena:2011jn, Maldacena:2012sf}.
 Traditionally,  most of the focus has been on bootstrapping correlators in position space,  mainly scalar four-point functions. However,   bootstrapping higher-point scalar and spinning correlators has been relatively less explored due to technical difficulties. One possible way to proceed in this direction is to find the right set of kinematic variables to aid bootstrapping.  Apart from the traditional position space approach,  there has been progress in momentum,  spinor helicity ~\cite{Maldacena:2011nz, McFadden:2011kk,  Coriano:2013jba,  Bzowski:2013sza,  Ghosh:2014kba,  Bzowski:2015pba,  Bzowski:2017poo,  Bzowski:2018fql,  Farrow:2018yni,  Isono:2019ihz,  Bautista:2019qxj,  Gillioz:2019lgs,  Baumann:2019oyu,  Baumann:2020dch,  Jain:2020rmw,  Jain:2020puw,  Jain:2021wyn,  Jain:2021qcl,  Jain:2021vrv,  Baumann:2021fxj,  Jain:2021gwa,  Jain:2021whr,  Gillioz:2022yze,  Marotta:2022jrp,  Jain:2023idr,  Bzowski:2023jwt,  S:2024zqp,  Jain:2024bza,  Aharony:2024nqs,  Marotta:2024sce,  Coriano:2024ssu,  Gillioz:2025yfb,  S:2025pmh} and Mellin variables in obtaining complementary results for CFT correlators. Recently,   there has been development in the form of a Twistor space formalism for three-dimensional Lorentzian CFT ~\cite{Baumann:2024ttn,  Bala:2025gmz,  Bala:2025jbh,  Bala:2025qxr,  Rost:2025uyj,  S:2025pmh,  Mazumdar:2025egx,  CarrilloGonzalez:2025qjk, Ansari:2025fvi}. Twistor space is a very promising approach,   as the action of the conformal generators and conservation is manifest. The twistor space approach has also been extended to accommodate non-conserved current correlators~\cite{Bala:2025qxr,   CarrilloGonzalez:2025qjk},   to incorporate supersymmetry~\cite{Bala:2025jbh,  Bala:2025qxr,  Mazumdar:2025egx}, and also to obtain Euclidean correlators which also obey global Ward-Takahashi identities~\cite{CarrilloGonzalez:2025qjk} up to three points.
 First steps towards twistor space 
 Four-point Wightman functions corresponding to special kinematics were taken in ~\cite{Ansari:2025fvi}. However,   the situation corresponding to general kinematics is much more complicated. Further work in this direction is required to explore the full potential of Twistor space. 

In order to look further for the best choice of the kinematic space,   another possibility is the Grassmannian space,   which is the space of planes that contains the kinematic data of the correlators. Recently,   the authors of ~\cite{Arundine:2026fbr} developed an orthogonal Grassmannian framework for CFT$_3$,   where correlators (or discontinuities thereof) were constrained via delta functions in the Grassmannian integral,   rendering the action of conformal generators algebraic and automatic. Four-point functions were constrained using physical principles such as unitarity,   which they show is extremely easy to implement in the Grassmannian.
 This type of development is quite reminiscent of the development in the modern study of scattering amplitudes ~\cite{Elvang:2013cua}. The Twistor variables used by Witten~\cite{Witten:2003nn} revealed that scattering amplitudes in $\mathcal{N}=4$ SYM localize on curves in Twistor Space. Later,   Twistor space and momentum Twistor space unveiled hidden geometric structures such as Grassmannian geometries~\cite{Arkani-Hamed:2012zlh} that make manifest dual conformal and Yangian symmetries of $\mathcal{N}=4$ SYM scattering amplitudes. In bootstrapping scattering amplitudes,   the introduction of the Grassmannian along with supersymmetry took the subject in an illuminating geometric direction. 
 
 This motivates us to find the supersymmetric extension of the CFT$_3$ Grassmannian,   and thus we focus on the class of CFT's which are supersymmetric. In SCFTs,   the bosonic and fermionic degrees of freedom are packaged together into multiplets,   and the supersymmetry generator allows us to determine relations between various component correlators constructed out of the superfields. Thus,   we set out to develop the supersymmetric Grassmannian to bootstrap the SCFT correlators,   which would give us a set of component correlators at once. 


The paper is organized as follows. In \textit{section \ref{sec:settingStage}},   we first set up the notation and conventions and also review the construction of the supersymmetric half integer spin multiplets. In \textit{section \ref{sec:Formalism}},   we derive a superspace extension of the orthogonal Grassmannian and obtain the superorthogonal Grassmannian that contains additional constraints on the orthogonal Grassmannian, which makes manifest the action of the super-conformal algebra OSp$(1|4)$.  In \textit{section \ref{sec:ExamplesOfSUCO}}, we use our formalism to reproduce known two- and three-point answers~\cite{Jain:2023idr} as well as extend the formalism to holographic four-point super-correlators. As an application, in  \textit{section \ref{sec:AdS4SYM}} we use it to bootstrap four-point $\mathcal{N}=1$ MHV super-correlators in AdS, and establish the connection between the gluon and gluino correlator. Further, we also establish the flat space limit directly in super-space and match with the results of ~\cite{Elvang:2011fx}. Our conclusions and possible future directions are presented in  \textit{section \ref{sec:Discussion}}.

The paper is supplemented with eight essential appendices. In \textit{Appendix \ref{app:Grassmanntwistorconstruction}},   we review the \(\mathcal{N}=1\) momentum superspace,   super-spinor helicity,   and Grassmann Twistor variables construction. In \textit{Appendix \ref{app:GeometryOfGrassmannian}},   we discuss the details of orthogonal Grassmannian notation,   branches,   and minor relations. In \textit{Appendix \ref{app:NewDeltaFunction}},   we review the properties of the Grassmann delta function appearing in \eqref{SuperGrassmannianNeq1}. We discuss another Grassmann bilinear solution to the supersymmetric Ward identities in Appendix \ref{app:bilinearUblock}. This is followed by appendices \ref{app:Neq1Integerspins} and \ref{app:scalar} which discuss the extension of our formalism to integer spin-currents and scalars respectively. In \textit{Appendix \ref{app:SUSYinSpinorHelicity}},   we review the existing supercorrelator construction in spinor helicity variables,   and we demonstrate the advantage Grassmannian formalism provides over it. In \textit{Appendix \ref{app:NkMHVSYM}}, we present some more results in different helicities that are not presented in the main text.\\
\textbf{Note}:~\textit{In our companion paper \cite{Bala:2026bdx}, we discuss the extension of our formalism to theories with $\mathcal{N}=2,3,4$ supersymmetry.}\\
\textbf{Note}: \textit{After the completion of this paper, we became aware of the work by Yu-Tin Huang, Chia-Kai Kuo, Yohan Liu, and  Jiajie Mei, which should appear concurrently \cite{Huang:2026tsh}}.

\section{Setting the (super-)stage for $\mathcal{N}=1$ SCFT$_3$}\label{sec:settingStage}
In this section,   we set up the essential notation, conventions, and review to construct the super-Grassmannian. 
To start with,   let us consider a conserved spin-$s$ symmetric traceless current $J_s$. In three dimensions,   the two independent components of $J_s$ are $J_s^{\pm}$,   which describe the positive and negative helicity currents.

In $\mathcal{N}=1$ supersymmetry,   we consider a conserved spin-$s$ symmetric traceless super-current $\mathbf{J}_s^{\pm}$ that describes the positive and negative helicity super-currents,   which contain spin-$s$ and spin-$(s+\frac{1}{2})$ currents as components. We describe these quantities in the language of super-spinor helicity and Grassmann twistor variables ~\cite{Jain:2023idr}\footnote{We review this construction in the appendix \ref{app:Grassmanntwistorconstruction}.}. For illustrative purposes,  we focus in the main-text, on half-integer spin super-currents, in particular the non-abelian spin-$\frac{1}{2}$ supermultiplet. We present the analogous construction for integer spins and scalars in the appendices \ref{app:scalar}, \ref{app:Neq1Integerspins}. The half-integer super-currents are given by,  
\begin{align}
    \mathbf{J}_s^{+}(\lambda,  \Bar{\lambda},  \xi)=J_s^{+}(\lambda,  \Bar{\lambda})+\frac{\xi}{\sqrt{2}}J_{s+\frac{1}{2}}^{+}(\lambda,  \Bar{\lambda}),  ~~\mathbf{J}_s^{-}(\lambda,  \Bar{\lambda},  \Bar{\xi})=J_s^{-}(\lambda,  \Bar{\lambda})+\frac{\Bar{\xi}}{\sqrt{2}}J_{s+\frac{1}{2}}^{-}(\lambda,  \Bar{\lambda}),  ~s\in\frac{2\mathbb{Z}_{\ge 0}+1}{2},  
\end{align}
$\xi$ and $\Bar{\xi}$ are Grassmann Fourier conjugate to each other,   and one can go from one description to the other by the Grassmann Fourier transform\footnote{The explicit Grassmann Fourier transform that takes us from one description to the other is,     
\begin{align}\label{xitoxibar}
    f(\Bar{\xi})=\int d\xi~ e^{\xi \Bar{\xi}}f(\xi)\iff f({\xi})=\int d\Bar{\xi}~e^{-\xi\Bar{\xi}}f(\bar\xi).
\end{align}}.  However,   we stick to the convention that for half-integer spin super-currents,   positive helicity currents are functions of $\xi$,   whereas negative helicity super-currents are functions of $\Bar{\xi}$. This is a choice that will simplify the results to follow\footnote{For integer spin super-currents,   on the other hand,   in appendix \ref{app:Neq1Integerspins},   we make the opposite choice,   yet again with some hindsight on what choice renders simpler expressions for correlators.}. 

The action of the supersymmetry generator acting on the super-currents in these variables is,  
\begin{align} \label{Qgenerator}
    \mathcal{Q}_a(\xi)=\frac{1}{\sqrt{2}}\bigg(\Bar{\lambda}_a \xi+\lambda_a\frac{\partial}{\partial \xi}\bigg),  ~\mathcal{Q}_a(\Bar{\xi})=\frac{1}{\sqrt{2}}\bigg(\Bar{\lambda}_a\frac{\partial}{\partial\Bar{\xi}}+\lambda_a \Bar{\xi}\bigg).
\end{align}
The special super-conformal generator,   on the other hand,   is given by,  
\begin{align}\label{Sgen}
    \mathcal{S}_a(\xi)=\frac{1}{\sqrt{2}}\bigg(\xi\frac{\partial}{\partial\lambda^a}+\frac{\partial}{\partial\xi}\frac{\partial}{\partial\Bar{\lambda}^a}\bigg),  ~ \mathcal{S}_a(\Bar{\xi})=\frac{1}{\sqrt{2}}\bigg(\frac{\partial}{\partial\Bar{\xi}}\frac{\partial}{\partial\lambda^a}+\Bar{\xi}\frac{\partial}{\partial\lambda^a}\bigg).
\end{align}
The representation of the conformal $\mathfrak{conf}(2,  1)$ generators acting on the super-currents is as follows:
\begin{align}\label{conformalGenerators}
\mathcal{P}_{ab} 
&= \lambda_{(a}\bar{\lambda}_{b)},   
& \mathcal{K}_{ab} 
&= \frac{\partial^2}{\partial \lambda^{(a}\,  \partial \bar{\lambda}^{b)}},  
\notag\\[6pt]
\mathcal{M}_{ab} 
&= \frac{1}{2}
\left(
    \lambda_{(a}\frac{\partial}{\partial \lambda^{b)}}
    + 
    \bar{\lambda}_{(a}\frac{\partial}{\partial \bar{\lambda}^{b)}}
\right),  
&
\mathcal{D} 
&= \frac{1}{2}
\left(
    \lambda^{a}\frac{\partial}{\partial \lambda^{a}}
    +
    \bar{\lambda}^{a}\frac{\partial}{\partial \bar{\lambda}^{a}}
    + 2
\right).
\end{align}

Note that the super-currents are having a particular weight under projective rescaling, in other words,   they are the eigenstates of the helicity generator\footnote{We will stick to the helicity generator action on the spin-half integer multiplet as defined. Similarly,   it can be written for an integer multiplet with the appropriate convention of Grassmann coordinates to be used.} given by,  
 \begin{align}
    &\qquad {h} (\mathbf{J}_s^{+}) = \frac{1}{2}\bigg( \bar\lambda^a\frac{\partial}{\partial\bar\lambda^a} - \lambda^a\frac{\partial}{\partial\lambda^a} +\xi\frac{\partial}{\partial\xi}  \bigg)(\mathbf{J}_s^{+}) = + s(\mathbf{J}_s^{+}) \notag \\ 
    &\qquad {h} (\mathbf{J}_s^{-}) = \frac{1}{2}\bigg( \bar\lambda^a\frac{\partial}{\partial\bar\lambda^a} - \lambda^a\frac{\partial}{\partial\lambda^a} -\bar\xi\frac{\partial}{\partial\bar\xi}  \bigg)(\mathbf{J}_s^{-}) = - s(\mathbf{J}_s^{-})
\end{align}

The objects of interest to us are correlation functions of the super-currents,  
\begin{align}
    \mathbf\Psi_n^{h_1\cdots h_n}=\langle \mathbf{J}_{s_1}^{h_1}\cdots \mathbf{J}_{s_n}^{h_n}\rangle.
\end{align}

Now the question is,   what correlator do we want to compute? There are many types of correlators we can consider,   such as time-ordered,   anti-time-ordered,   Wightman,   etc.  Time-ordered correlators or their Euclidean counterparts satisfy,  
\begin{align}
    \mathcal{K}_{\mu}\langle J_{s_1}\cdots J_{s_n}\rangle\propto \text{Current conservation Ward-Takahashi Identity}.
\end{align}
We focus on correlators that are identically conserved and obey a homogeneous special conformal Ward identity. In particular,   we focus on the discontinuity with respect to $p_1^2,   p_3^2$  of the time-ordered/Euclidean correlator that were considered in ~\cite{Arundine:2026fbr},   which have zero Ward-Takahashi identity.

Thus,   to obtain the homogeneous correlators,   we need to solve,  
\begin{align}
    \sum_{i=1}^{n}\mathcal{G}_i\mathbf\Psi_n^{h_1\cdots h_n}=0,  \mathcal{G}_i\in\{\mathcal{P}_{iab},  \mathcal{K}_{iab},  \mathcal{M}_{iab},  \mathcal{D}_{i},  h_i,  \mathcal{Q}_{ia},  \mathcal{S}_{ia}\}.
\end{align}
This will form the subject of the next section.

\section{Formalism: The $\mathcal{N}=1$ orthogonal super-Grassmannian}\label{sec:Formalism}
The idea of the Grassmannian representation is to make manifest as many of the symmetries of a correlator as possible. In ~\cite{Arundine:2026fbr},   the authors showed that the orthogonal Grassmannian $\text{OGr}(n,  2n)$ integral,  
\begin{align}\label{nonsusyGrassmannian}
    \psi_n^{h_1,  \cdots,   h_n}=\int \frac{d^{n\times 2n }C}{\text{Vol}(\mathbb{GL}(n))}\delta(C\cdot Q\cdot C^T)\delta(C\cdot \Lambda)F^{h_1\cdots h_n}(C),  
\end{align}
solves the translation as well as special conformal Ward identities automatically by virtue of the delta functions in the integral. 
Here,   $C$ is a $n\times 2n$ matrix with a $\mathbb{GL}(n)$ redundancy and $Q$ is the metric on the space $\mathbb{R}^{n,  n}$,  
\begin{align}\label{metric}
    Q=\begin{pmatrix}
        0_{n\times n}&\mathbb{I}_{n\times n}\\
        \mathbb{I}_{n\times n}&0_{n\times n}
    \end{pmatrix}.
\end{align}
$\Lambda$ is a $2n\times 2$ vector formed out of the spinor helicity variables of all the operators:
\begin{align}\label{Lambda}
    \Lambda=\begin{pmatrix}
        \lambda_1^1&\lambda_1^2\\
        \vdots &\vdots\\
        \lambda_n^1&\lambda_n^2\\
        \Bar{\lambda}_1^1&\Bar{\lambda}_1^2\\
        \vdots&\vdots\\
        \Bar{\lambda}_n^1&\Bar{\lambda}_n^2
    \end{pmatrix}.
\end{align}
Finally,   $f^{h_1\cdots h_n}(C)$ is a function of the $n\times n$ minors of the matrix $C$ which under $C\to G~ C,  G\in \mathbb{GL}(n)$ transforms with $\text{Det}(G)^{-(n+3)}$ to make the integrand $\mathbb{GL}$$(n)$ invariant.

In the supersymmetric case of interest to us,   we want an analogous super-Grassmannian integral representation that also makes the action of the supercharge $\mathcal{Q}$ and the superconformal generator $\mathcal{S}$ manifest. We will tackle this problem in two ways. First,   we write down the simplest possible $\mathbb{GL}(n)$ covariant extension of the ordinary Grassmannian by including the superspace coordinates.  In another approach,   we solve the Ward identities associated with the supercharges,   using which we constrain the possible form of the supersymmetric Grassmannian. As we shall see,   these two perspectives lead to the same result.
\subsection{A $\mathbb{GL}(n)$ covariant Grassmann extension of the Grassmannian}\label{subsec:superGrassmannExtension}
The central ingredient to construct a super-Grassmannian integral representation generalising \eqref{nonsusyGrassmannian} is the set of $2^{n}$ different $2n\times 1$ Grassmann ``phase" space vectors constructed out of the Grassmann variables $\xi$ and $\Bar{\xi}$.
\begin{align}\label{XiVectorNeq1}
    \Xi^{++\cdots +}=\begin{pmatrix}
        \xi_1\\
        \xi_2\\
        \vdots\\
        \xi_n\\
        \frac{\partial}{\partial\xi_1}\\
\frac{\partial}{\partial\xi_2}\\
        \vdots\\
        \frac{\partial}{\partial \xi_n}
    \end{pmatrix},  \Xi^{-+\cdots +}=\begin{pmatrix}
        \frac{\partial}{\partial\Bar{\xi}_1}\\
        \xi_2\\
        \vdots\\
        \xi_n\\
        \Bar{\xi}_1\\
\frac{\partial}{\partial \xi_2}\\
        \vdots\\
        \frac{\partial}{\partial \xi_n}
    \end{pmatrix},  \cdots,  \Xi^{-\cdots -}=\begin{pmatrix}
        \frac{\partial}{\partial\Bar{\xi}_1}\\
        \frac{\partial}{\partial\Bar{\xi}_2}\\
        \vdots\\
        \frac{\partial}{\partial\Bar{\xi}_n}\\
        \Bar{\xi}_1\\
        \Bar{\xi}_2\\
        \vdots\\
        \Bar{\xi}_n
    \end{pmatrix}.
\end{align}
Collectively,   we denote these vectors as $\Xi^{h_1\cdots h_n}$ to indicate which one we should use in a given helicity of a half-integer spin super-current.
These quantities can be thought of as the set of possible ``fermionic" square roots of the metric $Q$ since,  
\begin{align}
    \{\Xi^{A,  h_1\cdots h_n},  \Xi^{B,  h_1\cdots h_n}\}=Q^{AB}~~(A,  B=1,  \cdots 2n),  
\end{align}
by virtue of the anti-commutation relation $\{\xi_i,  \frac{\partial}{\partial \xi_j}\}=\delta_{ij}$. 
 With these objects in hand,   we are ready to define the super-Grassmannian for super-correlators involving half-integer super-currents $\mathcal{N}=1$ theories \footnote{There is another super-conformal building block that independently solves the super-conformal generators, see appendix (\ref{app:bilinearUblock}). However, the other building block will have a trivial contribution for $\mathcal{N}=1$ half-integer multiplets.}.
\begin{center}
\fbox{%
\begin{minipage}{0.99\textwidth}
\textbf{The $\mathcal{N}=1$ Grassmannian}
\begin{align}\label{SuperGrassmannianNeq1}
   &\mathbf{\Psi}_n^{h_1\cdots h_n}=\int \frac{d^{n\times 2n}C}{\text{Vol}(\mathbb{GL}(n))}\delta(C\cdot Q\cdot C^T)\delta(C\cdot\Lambda)\hat{\delta}(C\cdot\Xi^{h_1\cdots h_n})\mathcal{F}^{h_1\cdots h_n}(C),  
\end{align}
$\mathbf{\Psi}_n^{h_1\cdots h_n}$ is an n-point super-correlator involving half-integer spin super-currents. $\mathcal{F}^{h_1\cdots h_n}$ is a function of the $n\times n$ minors of $C$ and transforms with a factor of $\text{Det}(G)^{-(n-3)-1}$ under a $\mathbb{GL}(n)$ transformation. The above Grassmannian integral that incorporates the Grassmann delta function trivializes all $14$ super-conformal Ward identities.
\end{minipage}%
}
\end{center}
Let us motivate the above construction. Any modification of the non-supersymmetric Grassmannian integral \eqref{nonsusyGrassmannian} must respect the $\mathbb{GL}(n)$ redundancy. To connect the Grassmann phase space vector $\Xi^{h_1\cdots h_n}$ with $C$,   a natural choice is to contract them to form $C\cdot \Xi=C_{iA}(\Xi^{A,  h_1\cdots h_n}),   A=1,  \cdots,   2n,  i=1,  \cdots n$ which is a $1\times n$ Grassmann valued (including Grassmann derivatives) vector. To form a scalar quantity,   we can use SL$(n)$ invariant Levi-Civita symbol  as follows 
\begin{align}\label{deltaCXi1}
    \hat{\delta}(C\cdot \Xi^{h_1\cdots h_n})&=\frac{\epsilon_{i_1\cdots i_n}}{n!}(C\cdot\Xi^{h_1\cdots h_n})^{i_1}\cdots (C\cdot\Xi^{h_1\cdots h_n})^{i_n}\notag\\&=\int d^n \theta~e^{\theta\cdot C\cdot \Xi^{h_1\cdots h_n}}.
\end{align} 
This delta function is an operator (emphasized by the hat) due to the derivatives in $\Xi^{h_1\cdots h_n}$, which acts on what is on its right (which is $1$ from the perspective of the Grassmann derivatives). Its properties have been discussed in more detail in appendix \ref{app:NewDeltaFunction}.
This quantity is  $\mathbb{GL}(n)$ covariant as we will now check. Consider a $\mathbb{GL}(n)$ transformation $C_{iA}\to G_i^j C_{jA}$. The exponent transforms as $\theta^i C_{iA} (\Xi^{h_1\cdots h_n})^A\to \theta^i G_i^j C_{jA} (\Xi^{h_1\cdots h_n})^A$. We reabsorb this matrix into $\theta^i$ by defining $\tilde{\theta}^j=G^j_i\theta^i$. The measure is thus modified to $d^n\theta=\text{Det}(G)d^n\tilde{\theta}$, showing that this quantity transforms homogeneously under a $\mathbb{GL}(n)$ transformation.

We will discuss the details of evaluating this quantity with examples soon. Finally,   the properties that $\mathcal{F}^{h_1\cdots h_n}$ needs to have in order to ensure that the integrand is $\mathbb{GL}(n)$ invariant and the correlator has the correct helicity with respect to each super-current are the following:
\begin{align}\label{GLnEQ}
    \mathcal{F}^{h_1\cdots h_n}(GC)=\text{Det}(G)^{-(n-3)-1}\mathcal{F}^{h_1\cdots h_n}(C),  ~G\in \mathbb{GL}(n),  
\end{align}
and,  
\begin{align}\label{helicityEQ}
    \mathcal{F}^{h_1\cdots h_n}(\rho\cdot C)= \frac{1}{\rho_1^{2h_1}\cdots\rho_n^{2h_n}}\mathcal{F}^{h_1\cdots h_n}(C),  
\end{align}
where $\rho=\text{diag}(\rho_1,  \rho_2,  \cdots \rho_n,  \frac{1}{\rho_1},  \frac{1}{\rho_2},  \cdots,  \frac{1}{\rho_n})$ represents the independent little group scaling of each operator that is translated to the $C$ matrix from the scaling $\Lambda\to \rho\cdot \Lambda$ under a little group transformation $\{\lambda_1,  \cdots,  \lambda_n,  \Bar{\lambda}_1,  \cdots\Bar{\lambda}_n\}\to \{\rho_1\lambda_1,  \cdots,  \rho_n\lambda_n,  \frac{\Bar{\lambda}_1}{\rho_1},  \cdots,  \frac{\Bar{\lambda}_n}{\rho_n}\}$. Essentially,   the barred columns (first $n$ columns of the $C$ matrix) scale like $\Bar{\lambda}_i$ whereas the unbarred columns scale like $\lambda_i$. See appendix \ref{app:GeometryOfGrassmannian} for a detailed account of the construction and properties of the orthogonal Grassmannian.
\subsection{The super-Grassmannian from the supercharge Ward identities}\label{subsec:susyfromgenerator}
In this section, we derive the supersymmetric extension to the Orthogonal Grassmannian by solving the superconformal Ward identities. For that,   we start with the action of $\mathcal{S}$ along with $\mathcal{K},  \mathcal{P},  \mathcal{M}$ to fix the supercorrelator. We then show that the action of $\mathcal{Q}$ is also trivial on it.



The action of \(\mathcal{S}\) generator on the supercorrelator $\mathbf\Psi_n^{h_1\cdots h_n}$($\Lambda,  \Xi$) gives the following Ward identity,  
     \begin{align}\label{Saction1}
         \sum_i \mathcal{S}_{(i)a} \mathbf\Psi_n^{h_1\cdots h_n}(\Lambda,   \Xi^{h_1\cdots h_n})= 
         (\Xi^{T})^{h_1\cdots h_n(A)} \cdot \frac{\partial}{\partial\Lambda^{(A)a}} \mathbf\Psi_n^{h_1\cdots h_n}(\Lambda,   \Xi^{h_1\cdots h_n})=0.
     \end{align}
Let us consider a function of a linear combination of $\Lambda$ to be the solution to the Ward identity. We denote it with $\mathbf\Phi^{h_1\cdots h_n}_n(C\cdot\Lambda,  \Xi,  C)$ by introducing a real matrix $C$. Then the S ward identity implies
     \begin{align}\label{totalS}
         (\Xi^{T})^{h_1\cdots h_n(A)} \frac{\partial}{\partial\Lambda^{(A)a}} (\mathbf\Phi_n^{h_1\cdots h_n}(C\cdot\Lambda,   \Xi^{h_1\cdots h_n},  C))
         &= ((C\cdot\Xi)^T)^{h_1\cdots h_n(i)}  \frac{\partial}{\partial(C\cdot\Lambda)^{(i) a}} \mathbf\Phi_n^{h_1\cdots h_n}(C\cdot\Lambda,   \Xi^{h_1\cdots h_n},  C)\notag\\ 
        &=0 .
     \end{align}
For the above equation to be true for any arbitrary matrix C,   one's first guess would be,  
     \begin{align}\label{deltaCdotX}
\mathbf\Phi_n^{h_1\cdots h_n}(C\cdot\Lambda,  \Xi^{h_1\cdots h_n},  C)= \hat\delta(C\cdot\Xi^{h_1\cdots h_n}) f(C\cdot\Lambda,  C) ~~,  
     \end{align}
where the Grassmann delta function is defined in \eqref{deltaCXi1} as 
\begin{align}
    \hat{\delta}(C\cdot \Xi^{h_1\cdots h_n})  =\frac{\epsilon_{i_1\cdots i_n}}{n!}(C\cdot\Xi^{h_1\cdots h_n})^{i_1}\cdots (C\cdot\Xi^{h_1\cdots h_n})^{i_n}.
\end{align}
One can notice the remarkable property of this solution that the Grassmann part of the super correlator completely separates out. One advantage of this is,   if we apply the $\mathcal{K},   \mathcal{P}~\text{and}~\mathcal{M}$ generators on the above supercorrelator,   it will constrain the $f(C\cdot\Lambda,   C)$ analogous to steps done in ~\cite{Arundine:2026fbr} to be
       \begin{align}
f(C\cdot\Lambda,  C)=\delta(C\cdot\Lambda)\delta(C\cdot Q\cdot C^T) \mathcal{F}(C).
       \end{align}
Therefore,   an ansatz for the supercorrelator that satisfies \(\mathcal{S}\) would be\footnote{The factor $\hat\delta(C\cdot\Xi^{h_1\cdots h_n})$ in the $\mathbf\Psi^{h_1\cdots h_n}$ contains all the Grassmann coordinates, where $\Xi$ is also a Grassmann operator containing all the Grassmann coordinates. The only element that is left is the identity element `1' of the Grassmann algebra on which $\hat\delta(C\cdot\Xi)$ should act. } 
 \begin{align}\label{SUSYcorrelatoransatz}
\boxed{\mathbf\Phi_n^{h_1\cdots h_n}(C\cdot\Lambda,  \Xi^{h_1\cdots h_n},  C)= \hat\delta(C\cdot\Xi) \delta(C\cdot \Lambda)\delta(C\cdot Q\cdot C^T)\mathcal{F}(C)1} .
 \end{align}
    But there is a subtlety that we want to mention here. Looking at the above expression,   one would think that substituting it back in \eqref{totalS} would trivially give zero because of the form $x~\delta(x)$ as a distribution. But we should remember that $\hat\delta(C\cdot\Xi)$ is an operator, and one needs to be careful in handling operator equations of this form. So let us check by substituting the ansatz \eqref{SUSYcorrelatoransatz} in \eqref{totalS},   
    \begin{align}\label{Sonsupercorrelatoransatz}
    &((C\cdot\Xi)^T)^{h_1\cdots h_n(i)} \cdot \frac{\partial}{\partial(C\cdot\Lambda)^{i a}} \mathbf\Phi_n^{h_1\cdots h_n}(C\cdot\Lambda,   \Xi^{h_1\cdots h_n},  C) \notag\\
&=\delta(C\cdot Q\cdot C^T)\delta'(C\cdot\Lambda)_{a(i)}(C\cdot\Xi^{h_1\cdots h_n})^{(i)} \hat\delta^n(C\cdot\Xi)1
\end{align}
      Generically,   this is not zero,   but at the support of the SCT solving delta function \(\delta(C\cdot Q\cdot C^T)\),   this is 0,   since $C\cdot\Xi$ do indeed behave as if they were normal Grassmann variables, as we show in the appendix~\ref{app:NewDeltaFunction}. Therefore,   we have 
      \begin{equation}
          \delta(C\cdot Q\cdot C^T) (C\cdot \Xi) \hat \delta(C\cdot \Xi) = 0
      \end{equation}
      and the ansatz \eqref{SUSYcorrelatoransatz} indeed solves \(\mathcal{S}\).\\

With the supercorrelator that solves \(\mathcal{S}\),   one can show that the action of $\mathcal{Q}$ generator on the conformal supercorrelator $\mathbf{\Phi}_n^{h_1\cdots h_n}(C\cdot\Lambda,  C\cdot\Xi,  C)$ \eqref{SUSYcorrelatoransatz} is trivial for any arbitrary matrix C as follows: 
    \begin{align}
        \sum_i Q_{(i)a} \mathbf\Phi_n^{h_1\cdots h_n}(C\cdot\Lambda,  C\cdot\Xi,  C)&=\delta(C\cdot Q\cdot C^T )~((\Lambda^T)^A_a\cdot Q_{AB}\cdot \Xi^B )~\delta^n(C\cdot\Lambda)\delta^n(C\cdot \Xi)\mathcal{F(C)} \notag \\
        &=\delta(C\cdot Q\cdot C^T )~((((C\cdot Q)^T\cdot P)^T)^A_a\cdot Q_{AB}\cdot \Xi^B )~\delta^n(C\cdot\Lambda)\delta^n(C\cdot \Xi)\mathcal{F(C)} \notag \\
        &=\delta(C\cdot Q\cdot C^T )~((P^T)_a^{(i)} (C\cdot\Xi)_{(i)} )~\delta^n(C\cdot\Lambda)\delta^n(C\cdot \Xi)\mathcal{F(C)} \notag \\
        &= (P^T)_a^{(i)} \delta(C\cdot Q\cdot C^T )~\delta^n(C\cdot\Lambda)\cancelto{0}{(C\cdot\Xi)_{(i)}\delta^n(C\cdot \Xi)}\mathcal{F(C)} \notag \\
        &= 0
    \end{align}
where we have used $\delta(C\cdot\Lambda)$ to write $\Lambda$ in terms of C using an arbitrary basis vector $P_{n\times2}$ and then we have used the previous result i.e. $(C\cdot\Xi)\hat\delta(C\cdot\Xi)|_{\delta(C\cdot Q\cdot C^T)} = 0.$ \\
Therefore,   the general form of the supercorrelator $\mathbf\Psi_n^{h_1\cdots h_n}$ that not only solves \(P,  ~K,  \) and \(M\),   but also trivially solves \(\mathcal{Q}\) and \(\mathcal{S}\) is given by\footnote{There is another super-conformal building block that independently solves the superconformal generators refer to appendix (\ref{app:bilinearUblock}) for detail. However, the other building block will be having trivial contribution for $\mathcal{N}=1$ half-integer multiplet.}:
\begin{align}
    \boxed{\mathbf\Psi_n^{h_1\cdots h_n}= \int \frac{d^{n\times 2n}C}{\text{Vol}(\mathbb{GL}(n))}\delta^{\frac{n(n+1)}{2}}(C\cdot Q\cdot C^T)\delta^{n}(C\cdot\Lambda)\delta^n(C\cdot\Xi) \mathcal{F}(C).} 
\end{align}
where,   $\mathcal{F}$ has the same constraints as defined in section \ref{subsec:superGrassmannExtension} due to $D$ and helicity ($h$) generators. This is precisely what we also found earlier from demanding a GL$(n)$ covariant extension of the Grassmannian \eqref{SuperGrassmannianNeq1}. 
\section{Examples of Super-Correlation functions}\label{sec:ExamplesOfSUCO}
We now apply the formalism of the previous section to general conformal two-, three-, and four-point super-correlators.
\subsection{Two point functions}\label{subsec:twoPoint}
We now discuss the explicit construction of two-point functions of conserved currents. Two-point functions of non-identical operators are zero (modulo contact terms),   and thus we take identical operators with $h_1=h_2=h$. The $C$ matrix 
before gauge fixing is,  
\begin{align}
    C=\begin{pmatrix}
        c_{11}&c_{12}&c_{13}&c_{14}\\
        c_{21}&c_{22}&c_{23}&c_{24}
    \end{pmatrix}.
\end{align}
For the $(++)$ helicity configuration,   we have,  
\begin{align}
    \mathbf\Psi_2^{+s_1,  +s_2}=\int \frac{d^{2\times 4}C}{\text{Vol}(\mathbb{GL}(2))} \delta(C\cdot Q\cdot C^T)\delta(C\cdot \Lambda)\hat{\delta}(C\cdot\Xi^{++})\mathcal{F}^{+s_1,  +s_2}(C).
\end{align}
and to ensure the correct $\mathbb{GL}(2)$ covariance \eqref{GLnEQ} and little group weights \eqref{helicityEQ},   a natural choice is
\begin{align}
    \mathcal{F}^{+s_1,  +s_2}(C)=\frac{(\Bar{1}\Bar{2})^{2s}}{(1\Bar{1})^{2s}}
\end{align}
where $(\bar1\bar2)$ and $(1\bar1)$ are the minors of the C matrix as defined in appendix \ref{app:GeometryOfGrassmannian}.
The supersymmetric delta function using \eqref{deltaCXi1} equals,  
\begin{align}
    \hat{\delta}(C\cdot\Xi^{++})
    &=\frac{1}{2}\bigg(c_{13}c_{21}+c_{14}c_{22}-c_{11}c_{23}-c_{12}c_{24}\bigg)+(c_{11}c_{22}-c_{12}c_{21})\xi_1\xi_2\notag\\
    &=\frac{(1\Bar{1})+(2\Bar{2})+2\xi_1\xi_2(\Bar{1}\Bar{2})}{2}.
\end{align}
Thus, we find,  
\begin{align}
    \mathbf\Psi_2^{+s_1,  +s_2}=\frac{1}{2}\int \frac{d^{2\times 4}C}{\text{Vol}(\mathbb{GL}(2))}\delta(C\cdot Q\cdot C^T)\delta(C\cdot \Lambda)\bigg((1\Bar{1})+(2\Bar{2})+2\xi_1\xi_2(\Bar{1}\Bar{2})\bigg)\frac{(\Bar{1}\Bar{2})^{2s}}{(1\Bar{1})^{2s}}.
\end{align}
We now gauge fix and convert to spinor helicity variables. We choose,  
\begin{align}\label{2pointRightBranch}
    C=\begin{pmatrix}
        1&0&0&c_{12}\\
        0&1&-c_{12}&0
    \end{pmatrix}.
\end{align}
In this gauge,   $\delta(C\cdot\Lambda)=\delta^2(\lambda_1+c_{12}\Bar{\lambda}_2)\delta^2(\lambda_2-c_{12}\Bar{\lambda}_1)$. Three of the four delta functions combine to form $\delta^3(p_1^\mu+p_2^\mu)$ (up to Jacobian factors which are constants) whereas the fourth one localizes $c_{12}=-\frac{\langle 12\rangle}{E}$ where $E=2p_1=\langle\Bar{1}1\rangle$. Further we find $(1\Bar{1})=(2\Bar{2})=c_{12}=-\frac{\langle 12\rangle}{E}$ and $(\Bar{1}\Bar{2})=1$. This results in,  
\begin{align}
    \mathbf\Psi_2^{+s,  +s}=(-1)^{-2s}\bigg(-\frac{\langle 12\rangle}{E}+\xi_1\xi_2\bigg)\frac{E^{2(s+1)}}{\langle 12\rangle^{2(s+\frac{1}{2})}}=(-1)^{2s}\bigg(\xi_1\xi_2-\frac{\langle 12 \rangle}{E}\bigg)\frac{\langle \Bar{1} \Bar{2}\rangle^{2(s+\frac{1}{2})}}{E^{2s}},  
\end{align}
which is indeed the expected answer with the momentum-conserving delta function stripped off. A similar analysis can be carried out for the $(--)$ helicity with the $\mathcal{F}^{--}(C)$ given in the appendix \ref{app:NkMHVSYM}.
\subsection{Three point functions}\label{subsec:threePoint}
   Next,   we consider three-point functions,   focusing on the $(+++)$ and $(++-)$ helicities to illustrate our formalism. The half-integer spin multiplet three-point function leads to non-homogeneous correlators in Euclidean signature ~\cite{Jain:2023idr} that obey non-trivial Ward-Takahashi identities\footnote{From the bulk perspective,   these correspond to $\mathcal{N}=1$ Yang-Mills for $s_1=s_2=s_3=\frac{1}{2}$ and $\mathcal{N}=1$ Einstein super-gravity for $s_1=s_2=s_3=\frac{3}{2}$.}. However, we are bootstrapping certain discontinuities of these correlators, which are all homogeneous.
   \subsubsection*{$(+++)$ Helicity}
                Setting $n=3$ and $h_i=+s_i$ in our general solution \eqref{SuperGrassmannianNeq1} results in,  
                \begin{align}\label{JJJppps1s2s3}
                   \mathbf\Psi_3^{+s_1,  +s_2,  +s_3}= \int \frac{d^{3\times6}C}{Vol(\mathbb{GL}(3))} \delta(C^T\cdot Q \cdot C) \delta(C\cdot\Lambda)\hat\delta(C\cdot\Xi) \mathcal{F}^{+s_1,+s_2,+s_3}(C).
                \end{align}
 An ansatz that satisfies the $\mathbb{GL}(3)$ covariance property \eqref{GLnEQ} and helicity requirements \eqref{helicityEQ} is,  
       \begin{align}
         \boxed{ \mathcal{F}^{+s_1,+s_2,+s_3}(C) = \bigg(\frac{(\bar1\bar2\bar3)^{2s_1}(\bar1\bar21)^{2s_2-2s_1}(\bar1\bar31)^{2 s_3-2 s_1}  } {((1\bar12)(\bar23\bar3))^{s_2+s_3-s_1+\frac{1}{2}}}\bigg)} 
      \end{align}
Following the previous analysis of the two-point function,   one finds that the supersymmetric delta function using \eqref{deltaCXi1} equals,  
\begin{align}
    \hat{\delta}(C \cdot \Xi^{+++}) 
    &= \frac{1}{2}((\bar{1}\bar{2}2) + (\bar{1}\bar{3} 3))\xi_1 - \frac{1}{2}((\bar{1}\bar{2}1) - (\bar{2}\bar{3}3))\xi_2 \notag \\
    &\quad - \frac{1}{2}((\bar{1}\bar{3}1) + (\bar{2}\bar{3}2))\xi_3 - (\bar{1}\bar{2}\bar{3})\xi_1\xi_2\xi_3
\end{align}
To perform the integral in \eqref{JJJppps1s2s3},   we have to choose a branch as given in Appendix \ref{app:GeometryOfGrassmannian}. We choose the right branch $C$ matrix because $\mathcal{F}^{+++}$ and $\hat\delta(C\cdot\Xi)$ is zero in the left branch. So in a particular gauge on the right branch,  
                \begin{align}\label{3pointRightBranch}
    C=\begin{pmatrix}
        1&0&0&0&-c_{12}&-c_{13}\\
        0&1&0&c_{12}&0&-c_{23}\\
        0&0&1&c_{13}&c_{23}&0
    \end{pmatrix}.
\end{align}
The integral results in,  
\begin{align}
   \mathbf{\Psi}_3^{+s_1,  +s_2,  +s_3}& = 4  (-1)^{-s_1 -s_2 +s_3} \bigg(\frac{\langle\bar1\bar2\rangle}{E-2p_3}\bigg)^{\frac{1}{2}+s_1+s_2-s_3}\bigg(\frac{\langle\bar2\bar3\rangle}{E-2p_1}\bigg)^{\frac{1}{2}+s_2+s_3-s_1}\bigg(\frac{\langle\bar1\bar3\rangle}{E-2p_2}\bigg)^{\frac{1}{2}+s_1+s_3-s_2} \notag \\&~~~~~~~~~~~~~~~~~~~~~~~~~~~~~~~~~~~~~~~~~~~~~~~~~~~~~~~~~~~~~~\bigg(\xi_1\xi_2\xi_3-\frac{\langle23\rangle}{E}\xi_1-\frac{\langle31\rangle}{E}\xi_2-\frac{\langle12\rangle}{E}\xi_3\bigg) \notag\\
   &= 4 (-1)^{-s_1-s_2+s_3} \mathbf\Psi_3^{+(s_1+\frac{1}{2}),  +(s_2+\frac{1}{2}),  +2(s_3+\frac{1}{2})}~~ \mathbf{\Gamma_3}^{+++} ,  
\end{align}
where 
\begin{align}
\mathbf\Psi_3^{+(s_1+\frac{1}{2}),  +(s_2+\frac{1}{2}),  +2(s_3+\frac{1}{2})}=\bigg(\frac{\langle\bar1\bar2\rangle}{E-2p_3}\bigg)^{\frac{1}{2}+s_1+s_2-s_3}\bigg(\frac{\langle\bar2\bar3\rangle}{E-2p_1}\bigg)^{\frac{1}{2}+s_2+s_3-s_1}\bigg(\frac{\langle\bar1\bar3\rangle}{E-2p_2}\bigg)^{\frac{1}{2}+s_1+s_3-s_2}    
\end{align}
is the component correlator and $\Gamma_3^{+++}$ is the super-conformal block given by 
     \begin{align}
         \mathbf{\Gamma}_3^{+++}=\bigg(\xi_1\xi_2\xi_3-\frac{\langle23\rangle}{E}\xi_1-\frac{\langle31\rangle}{E}\xi_2-\frac{\langle12\rangle}{E}\xi_3\bigg) 
     \end{align}
which matches with the block as given in ~\cite{Jain:2023idr}\footnote{Note that the superconformal block given in the ~\cite{Jain:2023idr} matches with that of here upto a redefinition of the $\displaystyle\xi \xrightarrow{} \frac{1}{2\sqrt{2} }\xi^+$}.
   \subsubsection*{$(++-)$ Helicity}
                \begin{align}\label{JJJppms1s2s3}
                  \mathbf\Psi_3^{+s_1,+s_2,-s_3}= \int \frac{d^{3\times6}C}{Vol(\mathbb{GL}(3))} \delta(C^T\cdot Q \cdot C) \delta(C\cdot \Lambda)\hat\delta(C\cdot\Xi) \mathcal{F}^{+s_1,+s_2,-s_3}(C)
                \end{align}
    An ansatz that satisfies the $\mathbb{GL}(3)$ covariance property \eqref{GLnEQ} and helicity requirements \eqref{helicityEQ} is,  
       \begin{align}
         \boxed{ \mathcal{F}^{+s_1,+s_2,-s_3}(C) = \bigg(\frac{(\bar1\bar23)^{2s_1}(\bar1\bar21)^{2s_2-2s_1}(\bar131)^{2s_3-2s_1} } {((1\bar12)(\bar23\bar3))^{s_2+s_3-s_1+\frac{1}{2}}}\bigg)} 
      \end{align}
Similarly,   following the previous analysis,   the supersymmetric delta function using \eqref{deltaCXi1} equals,  
\begin{align}
    \hat{\delta}(C \cdot \Xi^{++-}) 
    &= \frac{1}{2}((\bar{1}\bar{2}2) - (\bar{1}\bar{3}3))\xi_1 - \frac{1}{2}((\bar{1}\bar{2}1) + (\bar{2}\bar{3}3))\xi_2 \notag \\
    &\quad + \frac{1}{2}((\bar{1}13) + (\bar{2}23))\bar{\xi}_3 - (\bar{1}\bar{2}3)\xi_1\xi_2\bar{\xi}_3
\end{align}

To perform the integral in \eqref{JJJppms1s2s3},   we choose the C matrix in the left branch,   where it has nontrivial support opposite to the case as in $+++$. So, in a particular choice of a gauge of the left branch,  
   \begin{align}\label{3pointLeftBranch}
    C=\begin{pmatrix}
        0&-c_{12}&-c_{13}&1&0&0\\
        c_{12}&0&-c_{23}&0&1&0\\
        c_{13}&c_{23}&0&0&0&1
    \end{pmatrix}.
\end{align}
 The integral results in,  
\begin{align}
    \mathbf\Psi_3^{+s_1,+s_2,-s_3}= &-4 (-1)^{2(s_2 - s_3)}  
\left( \frac{\langle \bar{1} \bar{2} \rangle}{E} \right)^{\frac{1}{2} + s_1 + s_2 - s_3}
\left( \frac{\langle \bar{2} 3 \rangle}{E - 2p_2} \right)^{\frac{1}{2} - s_1 + s_2 + s_3}
\left( \frac{\langle \bar{1} 3 \rangle}{E - 2p_1} \right)^{\frac{1}{2} + s_1 - s_2 + s_3} \notag \\
&~~~~~~~~~~~~~~~~~~~~~~~~~~~~~~~~~~~~~~~~\quad \times \left( \xi_1 \xi_2 \bar{\xi}_3 + \frac{E - 2p_2}{\langle \bar{2} 3 \rangle} \xi_1 - \frac{E - 2p_1}{\langle \bar{1} 3 \rangle} \xi_2 - \frac{E}{\langle \bar{1} \bar{2} \rangle} \bar{\xi}_3 \right)\notag\\
   &= -4 (-1)^{2(s_2-s_3)} \psi_3^{+(s_1+\frac{1}{2}),+(s_2+\frac{1}{2}),-(s_3+\frac{1}{2})}~~ \mathbf{\Gamma_3}^{+(s_1+\frac{1}{2}),+(s_2+\frac{1}{2}),-(s_3+\frac{1}{2})} ,  
\end{align}
where 
\begin{align}
\psi_3^{+(s_1+\frac{1}{2}),+(s_2+\frac{1}{2}),-(s_3+\frac{1}{2})}=\left( \frac{\langle \bar{1} \bar{2} \rangle}{E} \right)^{\frac{1}{2} + s_1 + s_2 - s_3}
\left( \frac{\langle \bar{2} 3 \rangle}{E - 2p_2} \right)^{\frac{1}{2} - s_1 + s_2 + s_3}
\left( \frac{\langle \bar{1} 3 \rangle}{E - 2p_1} \right)^{\frac{1}{2} + s_1 - s_2 + s_3}    
\end{align}
is the homogeneous component correlator and $\Gamma_3^{+++}$ is the super-conformal block given by 
     \begin{align}
         \mathbf{\Gamma}_3^{+(s_1+\frac{1}{2}),+(s_2+\frac{1}{2}),-(s_3+\frac{1}{2})}=\left( \xi_1 \xi_2 \bar{\xi}_3 + \frac{E - 2p_2}{\langle \bar{2} 3 \rangle} \xi_1 - \frac{E - 2p_1}{\langle \bar{1} 3 \rangle} \xi_2 - \frac{E}{\langle \bar{1} \bar{2} \rangle} \bar{\xi}_3 \right).
     \end{align}
Note that one can get other helicities $\mathcal{F}^{h_1h_2h_3}(C)$ from $\mathcal{F}^{+s_1,+s_2,+s_3}(C)$ by performing the helicity flipping operation given in appendix~\ref{app:NkMHVSYM} by switching the barred column of minors to unbarred and vice versa.

\subsection{Four point functions}\label{subsec:fourpointgen}  
Let us now set $n=4$ in our general expression \eqref{SuperGrassmannianNeq1}. The $\mathcal{N}=1$ supersymmetry will allow us to determine every component correlator including the top component viz $\langle J_{s_1+\frac{1}{2}}\cdots J_{s_4+\frac{1}{2}}\rangle$ from just the bottom component which is $\langle J_{s_1}\cdots J_{s_4}\rangle$. As we shall see,   the Grassmannian greatly simplifies these relations compared to the spinor helicity\footnote{In spinor-helicity variables, the relations among component correlators are not purely algebraic; instead, they take the form of first-order differential equations.
} approach, for more discussion see appendix \ref{app:SUSYinSpinorHelicity}. For concreteness,   we focus on the $(-+-+)$ helicity configuration where our Grassmann phase space vector is $\Xi^{-+-+}$. 

We begin by evaluating the Grassmann delta function,   $\hat{\delta}(C\cdot\Xi^{-+-+})$ in terms of the Grassmann variables $\Bar{\xi}_1,  \xi_2,  \Bar{\xi}_3,  \xi_4$ and the $4\times 4$ minors of the $C$ matrix,  
\begin{align}
    C=\begin{pmatrix}
        c_{11}\cdots c_{18}\\
        c_{21}\cdots c_{28}\\
        c_{31}\cdots c_{38}\\
        c_{41}\cdots c_{48}
    \end{pmatrix}.
\end{align}
We find,  
\begin{align}\label{GrassmannianDeltaExpansion4Point}
  &\hat{\delta}(C\cdot\Xi^{-+-+})\notag\\
  &=\frac{1}{4}\big((12\Bar{1}\Bar{2})+(34\Bar{3}\Bar{4})-(13\Bar{1}\Bar{3})-(24\Bar{2}\Bar{4})+(14\Bar{1}\Bar{4})+(23\Bar{2}\Bar{3})\big)\notag\\&+\frac{\Bar{\xi}_1\xi_2}{2}\big((13\Bar{2}\Bar{3})-(14\Bar{2}\Bar{4})\big)+\frac{\Bar{\xi}_1\Bar{\xi}_3}{2}\big((134\Bar{4})-(123\Bar{2})\big)+\frac{\Bar{\xi}_1\xi_4}{2}\big((12\Bar{2}\Bar{4})-(13\Bar{3}\Bar{4})\big)\notag\\
  &+\frac{\xi_2\Bar{\xi}_3}{2}\big((3 4\Bar{2}\Bar{4})-(13\Bar{1}\Bar{2})\big)+\frac{\xi_2\xi_4}{2}\big((3\Bar{2}\Bar{3}\Bar{4})-(1\Bar{1}\Bar{2}\Bar{4})\big)+\frac{\Bar{\xi}_3\xi_4}{2}\big((13\Bar{1}\Bar{4})-(23\Bar{2}\Bar{4})\big)\notag\\
  &-\Bar{\xi}_1\xi_2\Bar{\xi}_3\xi_4(13\Bar{2}\Bar{4}).
\end{align}
Identifying the Grassmannian Mandelstam variables defined in ~\cite{Arundine:2026fbr},  
\begin{align}
    S=(12\Bar{1}\Bar{2}),  \quad
    T=(14\Bar{1}\Bar{4}),  \quad
    U=(13\Bar{1}\Bar{3}),  
\end{align}
and using the following identities valid in the right branch of the orthogonal Grassmannian\footnote{One can observe that $\hat\delta(C\cdot\Xi^{-+-+})$ has a non-trivial support only at the right branch.}:
\begin{align}
    (12\Bar{1}\Bar{2})=(34\Bar{3}\Bar{4}),   \quad
    (14\Bar{1}\Bar{4})=(23\Bar{2}\Bar{3}),   \quad
    (13\Bar{1}\Bar{3})=(24\Bar{2}\Bar{4}).
\end{align}
We find,  
\begin{align}\label{psi4gen}
    \mathbf\Psi_4^{-+-+}=\int \frac{d^{4\times 8}C}{\text{Vol}(\mathbb{GL}(n))}&\delta(C\cdot Q\cdot C^T)\delta(C\cdot \Lambda)\Bigg(\frac{1}{2}\big(S+T-U\big)\notag\\&+\frac{\Bar{\xi}_1\xi_2}{2}\big((13\Bar{2}\Bar{3})-(14\Bar{2}\Bar{4})\big)+\frac{\Bar{\xi}_1\Bar{\xi}_3}{2}\big((134\Bar{4})-(123\Bar{2})\big)+\frac{\Bar{\xi}_1\xi_4}{2}\big((12\Bar{2}\Bar{4})-(13\Bar{3}\Bar{4})\big)\notag\\
  &+\frac{\xi_2\Bar{\xi}_3}{2}\big((3 4\Bar{2}\Bar{4})-(13\Bar{1}\Bar{2})\big)+\frac{\xi_2\xi_4}{2}\big((3\Bar{2}\Bar{3}\Bar{4})-(1\Bar{1}\Bar{2}\Bar{4})\big)+\frac{\Bar{\xi}_3\xi_4}{2}\big((13\Bar{1}\Bar{4})-(23\Bar{2}\Bar{4})\big)\notag\\
  &-\Bar{\xi}_1\xi_2\Bar{\xi}_3\xi_4(13\Bar{2}\Bar{4})\Bigg)\mathcal{F}_4^{-+-+}(C),  
\end{align}
There is a beautiful structure to the above equation such that every component correlator will be fixed just by determining the function $\mathcal{F}_4^{-+-+}(C)$. If we do the super-current expansion in the LHS of the above equation,   we get bottom and top component correlators as\footnote{$\displaystyle \int \mathcal{D}C \equiv \int \frac{d^{4\times8}C}{Vol(\mathbb{GL}(n))}\delta(C\cdot Q\cdot C^T)\delta(C\cdot\Lambda)$ }
       \begin{align}\label{Grassmaniancorrelatorrelation}
             \mathbf\Psi_4^{-s_1+s_2-s_3+s_4} &= \int \mathcal{D}C \frac{(S+T-U)}{2} \mathcal{F}_4^{-+-+}(C)=\int\mathcal{D}C A_{s_1,  s_2,  s_3,  s_4}(C) \notag\\
            ~~~\text{and}~~~ -\frac{\mathbf\Psi_4^{-(s_1+\frac{1}{2})+(s_2+\frac{1}{2})-(s_3+\frac{1}{2})+(s_4+\frac{1}{2})}}{4} &= \int \mathcal{D}C ~(13\bar2\bar4)~ \mathcal{F}_4^{-+-+}(C) =\int\mathcal{D}C A_{s_1+\frac{1}{2},  s_2+\frac{1}{2},  s_3+\frac{1}{2},  s_4+\frac{1}{4}}(C)
       \end{align}
With the structure of the above \eqref{Grassmaniancorrelatorrelation} we can obtain the relation between the component correlators at the level of the integrand as,  
\begin{align}\label{relatingTopAndBottomCorrelators}
    \mathcal{F}^{-+-+}(C)&=-\frac{1}{4(13\Bar{2}\Bar{4})}A_{s_1+\frac{1}{2},  s_2+\frac{1}{2},  s_3+\frac{1}{2},  s_4+\frac{1}{4}}(C)\notag\\
    &=\frac{2}{S+T-U}A_{s_1,  s_2,  s_3,  s_4}(C)\notag
\end{align}
\begin{align}
   \implies A_{s_1+\frac{1}{2},  s_2+\frac{1}{2},  s_3+\frac{1}{2},  s_4+\frac{1}{4}}(C)=-\frac{8(13\Bar{2}\Bar{4})}{S+T-U}A_{s_1,  s_2,  s_3,  s_4}(C),      
\end{align}
with similar relations relating every other component correlator to the bottom component\footnote{Let us note that, as discussed in \cite{Arundine:2026fbr}, these correlators exhibit discontinuities with respect to two external momenta. It should not be difficult to establish analogous relations among correlators without such discontinuities.},   showing us the utility of $\mathcal{N}=1$ supersymmetry in being able to construct seven correlators starting with just one. The fact that the constraint between the component correlator in Grassmannian space is algebraic as compared to differential constraints in the spinor helicity variables is really an advantage; details can be referred to in the appendix \ref{app:SUSYinSpinorHelicity}. Later,   we will use this algebraic relation in the AdS$_4$ case.

\subsection{Connection to super-Twistor space}\label{subsec:connectionToTwistor}
     In this section,   we make a connection between the Grassmannian $C$ matrix and the Schwinger parameters used to represent super-twistor correlators, using three-point functions to illustrate this. We consider the $(+++)$ three-point half-integer spin super-correlator as an example. We perform the half-Fourier transform from twistor space to spinor-helicity,   which will yield the gauge-fixed Grassmannian results. This correlator in the language of super-twistors $\mathcal{Z}=(\lambda_i^a,  \Bar{\mu}_{ia'},  \psi=e^{-\frac{i\pi}{4}}\xi)$ is given by~\cite{Bala:2025qxr}\footnote{The expression for this correlator in dual-twistor variables is ~\cite{Bala:2025jbh},  
     {\small
\begin{align}\label{3deltaSusydotNHhalfint}
    \langle0|\mathbf{\hat{J}_{s_1}^+ }(\mathcal{W}_1)\mathbf{\hat{J}_{s_2}^+ }(\mathcal{W}_2)\mathbf{\hat{J}_{s_3}^+ }(\mathcal{W}_3)|0\rangle =  (-i)^{s_T} \delta^{[s_1+s_2-s_3+\frac{\mathcal{N}}{2}]}(\mathcal{W}_1\cdot \mathcal{W}_2)\delta^{[s_2+s_3-s_1+\frac{\mathcal{N}}{2}]}(\mathcal{W}_2\cdot \mathcal{W}_3)\delta^{[s_3+s_1-s_2+\frac{\mathcal{N}}{2}]}(\mathcal{W}_3 \cdot \mathcal{W}_1).
\end{align}}
$\mathcal{W}_i=(\mu_{ia},  \Bar{\lambda}_i^{a'},  \Bar{\psi})$ are dual super-twistors. However,   this expression is Grassmann even,   whereas the natural Grassmannian for half-integer spin currents \eqref{SuperGrassmannianNeq1} is Grassmann odd,   which is naturally connected to the super-twistor expression \eqref{delta4+++}},  
\begin{align}\label{delta4+++}
    &\langle0|\mathbf{\hat{J}_{s_1}^+ }(\mathcal{Z}_1)\mathbf{\hat{J}_{s_2}^+ }(\mathcal{Z}_2)\mathbf{\hat{J}_{s_3}^+ }(\mathcal{Z}_3)|0\rangle\notag\\&= (-i)^{s_T}\delta^{[s_1+s_2+s_3]}(\mathcal{Z}_1\cdot\mathcal{Z}_2) \int dc_{23} dc_{31} c_{23}^{-s_2-s_3+s_1} c_{31}^{-s_3-s_1+s_2} \delta^{4|\mathcal{N}}(\mathcal{Z}_3^{\mathcal{A}}+ c_{31} \mathcal{Z}_2^{\mathcal{A}}+c_{23}\mathcal{Z}_1^{\mathcal{A}}).
\end{align}
We shall show how performing a half-Fourier transform to spinor helicity from \eqref{delta4+++} will result in the Schwinger parameter space expressions that correspond to a particular gauge-fixing choice of the Grassmannian $C$ matrix. The spinor helicity expression is given by,  
\begin{align}
      &\mathbf\Psi_3^{+++}=\prod_{i=1}^{3}\big(\int d^2\Bar{\mu}_i e^{-i\Bar{\lambda}_i\cdot\Bar{\mu}_i}\big) \langle0|\mathbf{\hat{J}_{s_1}^+ }(\mathcal{Z}_1)\mathbf{\hat{J}_{s_2}^+ }(\mathcal{Z}_2)\mathbf{\hat{J}_{s_3}^+ }(\mathcal{Z}_3)|0\rangle\notag \\&=\big(\int d^2\Bar{\mu}_i e^{-i\Bar{\lambda}_i\cdot\Bar{\mu}_i}\big) \int dc_{12}dc_{23}dc_{31}c_{12}^{s_1+s_2+s_3}c_{23}^{-s_2-s_3+s_1}c_{31}^{-s_3-s_1+s_2}e^{i c_{12}\mathcal{Z}_1\cdot\mathcal{Z}_2}\delta^{4|1}(\mathcal{Z}_3^{\mathcal{A}}+ c_{31} \mathcal{Z}_2^{\mathcal{A}}+c_{23}\mathcal{Z}_1^{\mathcal{A}}) \notag \\
      &=\int dc_{12}dc_{23}dc_{31}c_{12}^{s_1+s_2+s_3}c_{23}^{-s_2-s_3+s_1}c_{31}^{-s_3-s_1+s_2}\delta(C\cdot\Lambda)\bigg(\xi_3+c_{31}\xi_2+c_{23}\xi_1+c_{23}\xi_1\xi_2\xi_3\bigg),  
  \end{align}
where $C$ in the above equation is given by,  
\begin{align}
    C=\begin{pmatrix}\label{3pointrightbranchtwistor}
        c_{23}&c_{31}&1&0&0&0\\
        0&-c_{12}&0&1&0&-c_{23}\\
        c_{12}&0&0&0&1&-c_{31}
    \end{pmatrix}.
\end{align}
This $C$ matrix corresponds to choosing the right branch of the Grassmannian since we have $(1\bar 2\bar 3)=0$,   which is its defining condition. The C matrix in the above equation and in \eqref{3pointRightBranch} are related by a $\mathbb{GL}(3)$ transformation.\\

We will now specialize to AdS$_4$ $\mathcal{N}=1$ super Yang-Mills theory,   which corresponds to focusing on the spin-$\frac{1}{2}$ super-multiplet from a CFT perspective. 

\section{Application: AdS$_4$ Supersymmetric Yang-Mills theory}\label{sec:AdS4SYM}
The aim of this section is to apply the results of subsection \ref{subsec:fourpointgen} to $\mathcal{N}=1$ super Yang-Mills theory in AdS$_4$ with gauge group SU(N). In the bulk,   we have gluons and gluinos both in the adjoint representation of SU(N). Each of these fields has on-shell positive and negative helicity degrees of freedom matching with the boundary conserved operator in 3-d CFT. We will focus on the boundary-to-boundary supercorrelator of the gluon and gluino, which can be effectively represented in terms of the correlators of the  super-current:
\begin{align}\label{Neq1SYMAdSsuperfield}
    &\mathbf{J}_{\frac{1}{2}}^{+}=O_{\frac{1}{2}}^{+}+\frac{\xi}{\sqrt{2}}J^{+}\notag\\
    &\mathbf{J}_{\frac{1}{2}}^{-}=O_{\frac{1}{2}}^-+\frac{\Bar{\xi}}{\sqrt{2}}J^-,  
\end{align}
    where $O_{\frac{1}{2}}^{\pm}$ and $J^{\pm}$ represents gluino and gluon in the bulk respectively.

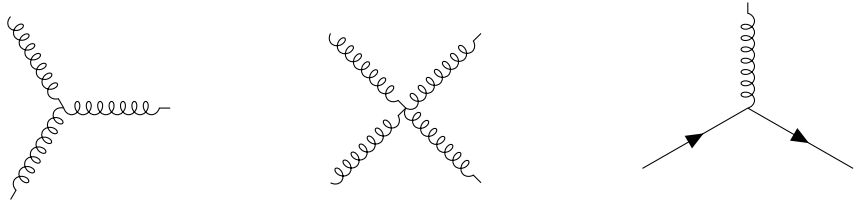
\begin{figure}[h]
\centering
\begin{tikzpicture}[baseline=(current bounding box.center)]

\begin{scope}[xshift=-4.5cm]
\begin{feynman}
\vertex (v);
\vertex at ($(v)+(120:1.4)$) (v1);
\vertex at ($(v)+(0:1.4)$)   (v2);
\vertex at ($(v)+(-120:1.4)$) (v3);

\diagram* {
(v1) -- [gluon] (v) -- [gluon] (v2),  
(v)  -- [gluon] (v3),  
};
\end{feynman}
\end{scope}

\begin{scope}[xshift=0cm]
\begin{feynman}
\vertex (v);
\vertex at ($(v)+(135:1.4)$) (v1);
\vertex at ($(v)+(45:1.4)$)  (v2);
\vertex at ($(v)+(-135:1.4)$) (v3);
\vertex at ($(v)+(-45:1.4)$)  (v4);

\diagram* {
(v1) -- [gluon] (v) -- [gluon] (v2),  
(v3) -- [gluon] (v) -- [gluon] (v4),  
};
\end{feynman}
\end{scope}

\begin{scope}[xshift=4.5cm]
\begin{feynman}
\vertex (v);
\vertex at ($(v)+(210:1.6)$) (f1);
\vertex at ($(v)+(330:1.6)$)   (f2);
\vertex at ($(v)+(90:1.4)$)  (g);

\diagram* {
(f1) -- [fermion] (v) -- [fermion] (f2),  
(v)  -- [gluon] (g),  
};
\end{feynman}
\end{scope}

\end{tikzpicture}
\caption{\textbf{The vertices in $\mathcal{N}=1$ SYM theory.}}\label{fig:Neq1SYM}
\end{figure}

$\mathcal{N}=1$ super Yang-Mills theory is classically conformally invariant, and its spectrum is the same in (A)dS$_4$ and flat spacetime. We have gluons and gluinos both in the adjoint representation of $SU(N)$. The supersymmetric action in the flat space and AdS will be similar modulo boundary terms because of them being conformally equivalent. So one can guess the allowed bulk interactions that are present in this theory are schematically shown in Figure \ref{fig:Neq1SYM}, see ~\cite{Freedman:2012zz}. We can immediately draw certain conclusions from this structure. For one,   the tree-level gluon four-point function is the same as in pure Yang-Mills theory. Further,   since supersymmetry connects the gluon and gluino four-point functions,   we can use the latter to determine the former. An interesting fact is that the gluino four-point function receives contributions from just gluon exchanges and has no contact diagram contribution,   unlike the gluon four-point function.

To begin with,   we stick to the helicity $-+-+$ and we find every other component correlator in terms of  the spin-$\frac{1}{2}$ Grassmannian correlator($A_{\frac{1}{2},  \frac{1}{2},  \frac{1}{2},  \frac{1}{2}}^{-+-+}$) in a similar approach that we followed to get \eqref{relatingTopAndBottomCorrelators} as follows,  
\begin{align}\label{susyconstraints}
    &\qquad \qquad\qquad\qquad\qquad\qquad A_{1,  1,  1,  1}^{-+-+}=\frac{8(1\Bar{2}3\Bar{4})}{S+T-U}A_{\frac{1}{2},  \frac{1}{2},  \frac{1}{2},  \frac{1}{2}}^{-+-+},  \notag\\
    &A_{1,  1,  \frac{1}{2},  \frac{1}{2}}^{-+-+}=\frac{2\big((1\Bar{2}4\Bar{4})-(1\Bar{2}3\Bar{3})\big)}{S+T-U}A_{\frac{1}{2},  \frac{1}{2},  \frac{1}{2},  \frac{1}{2}}^{-+-+},  \qquad\qquad A_{1,  \frac{1}{2},  1,  \frac{1}{2}}^{-+-+}=-\frac{2\big((134\Bar{4})-(123\Bar{2})\big)}{S+T-U}A_{\frac{1}{2},  \frac{1}{2},  \frac{1}{2},  \frac{1}{2}}^{-+-+},  \notag\\
    &A_{1,  \frac{1}{2},  \frac{1}{2},  1}^{-+-+}=\frac{2\big((12\Bar{2}\Bar{4})-(13\Bar{3}\Bar{4})\big)}{S+T-U}A_{\frac{1}{2},  \frac{1}{2},  \frac{1}{2},  \frac{1}{2}}^{-+-+},  \qquad\qquad  A_{\frac{1}{2},  1,  1,  \frac{1}{2}}^{-+-+}=-\frac{2\big((13\Bar{1}\Bar{2})-(3 4\Bar{2}\Bar{4})\big)}{S+T-U}A_{\frac{1}{2},  \frac{1}{2},  \frac{1}{2},  \frac{1}{2}}^{-+-+},  \notag\\
    &A_{\frac{1}{2},  1,  \frac{1}{2},  1}^{-+-+}=\frac{2\big((1\Bar{1}\Bar{2}\Bar{4})-(3\Bar{2}\Bar{3}\Bar{4})\big)}{S+T-U}A_{\frac{1}{2},  \frac{1}{2},  \frac{1}{2},  \frac{1}{2}}^{-+-+},  \qquad\qquad A_{\frac{1}{2},  \frac{1}{2},  1,  1}^{-+-+}=\frac{2\big((13\Bar{1}\Bar{4})-(23\Bar{2}\Bar{4})\big)}{S+T-U}A_{\frac{1}{2},  \frac{1}{2},  \frac{1}{2},  \frac{1}{2}}^{-+-+}.
\end{align}

In this section,   our aim is to bootstrap the color-ordered gluino four-point function and use supersymmetry to determine the gluon four-point function that was found in ~\cite{Arundine:2026fbr}. 
\subsection{Bootstrapping  The spin$-\frac{1}{2}$ gluino four point function}
We follow the bootstrapping principles set forth in ~\cite{Arundine:2026fbr}. We determine the (color-ordered) correlator by demanding that it factorizes consistently into a product of three-point functions at $S=0$ and $T=0$. We demand,  
\begin{align}
    \text{Res}_{S=0}\bigg(A^{-+-+}_{\frac{1}{2},  \frac{1}{2},  \frac{1}{2},  \frac{1}{2}}\bigg)=\sum_{h=-1,  +1}A_{\frac{1}{2},  \frac{1}{2}, 1}^{-+h}A_{1, \frac{1}{2}, \frac{1}{2}}^{(-h)-+}.
\end{align}
By using the three-point Grassmannian integral,  we can show that the following expressions lead to the correct spinor helicity correlators:
\begin{align}
    &A^{-+-}_{\frac{1}{2}, \frac{1}{2}, 1}=-\frac{(1 \Bar{2}I_s)}{(1\Bar{1}\Bar{I}_s)}, \qquad A_{1, \frac{1}{2}, \frac{1}{2}}^{+-+}=\frac{(\Bar{4}\Bar{I}_s3)}{(\Bar{3}3I_s)}, \notag\\
    &A_{\frac{1}{2}, \frac{1}{2}, 1}^{-++}=-\frac{(1\Bar{2}\Bar{I}_s)}{(1\Bar{1}I_s)}, \qquad A_{1, \frac{1}{2}, \frac{1}{2}}^{--+}=\frac{(\Bar{4}I_s3)}{(\Bar{3}3\Bar{I}_s)}.
\end{align}
Putting these together,  we find, 
\begin{align}
    \text{Res}_{S=0}\bigg(A^{-+-+}_{\frac{1}{2}, \frac{1}{2}, \frac{1}{2}, \frac{1}{2}}\bigg)=-\bigg(\frac{(1\Bar{2}I_s)(3\Bar{4}\Bar{I}_s)}{(1\Bar{1}\Bar{I}_s)(3\Bar{3}I_s)}+\frac{(1\Bar{2}\Bar{I}_s)(3\Bar{4}I_s)}{(1\Bar{1}I_s)(3\Bar{3}\Bar{I}_s)}\bigg).
\end{align}
We now make use of the following Pl\"ucker relations:
\begin{align}
    (1\Bar{2}I_s)=-\frac{(1\Bar{1}\Bar{I}_s)(\Bar{2}1I_s)^2}{(1\Bar{I}_sI_s)(\Bar{2}\Bar{I}_sI_s)}, \qquad(3\Bar{4}\Bar{I}_s)=-\frac{(3\Bar{3}I_s)(\Bar{4}3\Bar{I}_s)^2}{(3\Bar{I}_sI_s)(\Bar{4}\Bar{I}_sI_s)}, \notag \\ 
    ~ (1\Bar{2}\bar I_s)=-\frac{(1\Bar{1}{I}_s)(\Bar{2}1\bar I_s)^2}{(1{I}_s\bar I_s)(\Bar{2}{I}_s\bar I_s)}, \qquad(3\Bar{4}{I}_s)=-\frac{(3\Bar{3}\bar I_s)(\Bar{4}3{I}_s)^2}{(3{I}_s\bar I_s)(\Bar{4}{I}_s\bar I_s)}.
\end{align}
 Using these relations,  we find, 
\begin{align}
    \text{Res}_{S=0}\bigg(A^{-+-+}_{\frac{1}{2}, \frac{1}{2}, \frac{1}{2}, \frac{1}{2}}\bigg)&=\frac{1}{(1I_s\Bar{I}_s)(\Bar{2}I_s\Bar{I}_s)(3I_s\Bar{I}_s)(\Bar{4}I_s\Bar{I}_s)}\bigg((1 \Bar{2}I_s)^2(3\Bar{4}\Bar{I}_s)^2+(1\Bar{2}\Bar{I}_s)^2(3\Bar{4}I_s)^2\bigg) \notag \\
    & =\frac{1}{(1I_s\Bar{I}_s)(\Bar{2}I_s\Bar{I}_s)(3I_s\Bar{I}_s)(\Bar{4}I_s\Bar{I}_s)}\bigg((1 \Bar{2}I_s)(3\Bar{4}\Bar{I}_s)+(1\Bar{2}\Bar{I}_s)(3\Bar{4}I_s)\bigg)^2 \notag \\
    & =\frac{1}{-2(1\bar23\bar4)(T+U)}(1\bar23\bar4)^2=-\frac{(1\Bar{2}3\Bar{4})}{2(T+U)}
\end{align}
where we have used the relation $$2(1\Bar{2}I_s)(1\Bar{2}\Bar{I}_s)(3\Bar{4}\Bar{I}_s)(3\Bar{4}I_s)=0$$ in going from the first line to the second line in the last equation when completing the square.  Also,  we have used 
\begin{align}
    &(1\Bar{2}I_s)(3\Bar{4}\Bar{I}_s)+(1\Bar{2}\Bar{I}_s)(3\Bar{4}I_s)=(1\Bar{2}3\Bar{4}),  \notag \\ 
    \text{and} ~~~
    &(1I_s\Bar{I}_s)(\Bar{2}I_s\Bar{I}_s)(3 I_s\Bar{I}_s)(\Bar{4}I_s\Bar{I}_s)=-2(1\Bar{2}3\Bar{4})(T+U)  
\end{align} in going from the second to the third line. 
A similar analysis shows, 
\begin{align}
    \text{Res}_{T=0}\bigg(A^{-+-+}_{\frac{1}{2}, \frac{1}{2}, \frac{1}{2}, \frac{1}{2}}\bigg)=-\frac{(1\Bar{2}3\Bar{4})}{2(S+U)}.
\end{align}
One of the possible expressions for the gluino 4-point correlator by gluing the factorization limit that reproduces these residues is, 
\begin{align}\label{Ohfourpoint}
    A^{-+-+}_{\frac{1}{2}, \frac{1}{2}, \frac{1}{2}, \frac{1}{2}}=-\frac{(1\Bar{2}3\Bar{4})}{2(S+T+U)}\bigg(\frac{1}{S}+\frac{1}{T}\bigg).
\end{align}

One great exercise would be to take the above answer and compute its normal orthogonal Grassmanian integral to get the spinor helicity using the correct pole enclosure as given in the \cite{Arundine:2026fbr} for the gluon correlator, to get the spinor helicity correlators which are the discontinuity with respect to $p_1^2$ and $p_3^2$. This spinor helicity answer can be matched with the fermion correlators calculation\cite{Chen:2025foq,Chen:2025ljl} done in the AdS$_4$ with appropriate discontinuity.  As will be shown below,  the above answer produces the correct flat space limit as well as the correct spin-1 four-point function in AdS$_4$.

\subsection{Deriving the Gluon four-point function}
We now use \eqref{Ohfourpoint} in \eqref{susyconstraints} to determine the gluon four-point correlator. We obtain, 
\begin{align}
    A_{1, 1, 1, 1}^{-+-+}&=-\frac{8(1\Bar{2}3\Bar{4})}{S+T-U}A_{\frac{1}{2}, \frac{1}{2}, \frac{1}{2}, \frac{1}{2}}^{-+-+}=-\frac{4(1\Bar{2}3\Bar{4})^2}{(S+T-U)(S+T+U)}\bigg(\frac{1}{S}+\frac{1}{T}\bigg)\notag\\
    &=-2\bigg(\frac{1}{S+T+U}+\frac{1}{S+T-U}\bigg)\frac{(1\Bar{2}3\Bar{4})^2}{ST}, 
\end{align}
which is exactly the result found in ~\cite{Arundine:2026fbr} for the four point gluon correlator!\footnote{More precisely,  it's discontinuity with respect to $p_1^2$ and $p_3^2$.} Note that the spin$-\frac{1}{2}$ four-point function,  unlike the gluon four-point function,  has no pole of the form $\frac{1}{S+T-U}$. However,  supersymmetry reproduces this additional pole,  leading to the correct result. This exercise illustrates the power and simplicity of $\mathcal{N}=1$ supersymmetry while also serving as a check on the correctness of our formalism.
\subsection{The Flat Space Limit}\label{subsec:flatSpaceLimit}
We now make a connection to the flat space $\mathcal{N}=1$ super Yang-Mills theory scattering amplitude ~\cite{Elvang:2011fx}. To obtain the flat space scattering amplitude from the Grassmannian integral,  we first gauge-fix the $\mathbb{GL}(n)$ redundancy of the $C$ matrix. We are working in the right branch where we can choose $C$ to take the following form:
\begin{align}\label{rightBranchGaugeFixing}
    C=\begin{pmatrix}
        1&0&0&0&0&-c_{12}&-c_{13}&-c_{14}\\
        0&1&0&0&c_{12}&0&-c_{23}&-c_{24}\\
        0&0&1&0&c_{13}&c_{23}&0&-c_{34}\\
        0&0&0&1&c_{14}&c_{24}&c_{34}&0
    \end{pmatrix}.
\end{align}
We can then use $\delta(C\cdot\Lambda)$ to obtain the momentum-conserving delta function and solve for $5$ out of the $6$ Schwinger parameters $c_{ij}$. We can parametrize them as follows,  following ~\cite{Arundine:2026fbr}:
\begin{align}
    c_{ij}=\frac{\langle i j\rangle}{E}+\frac{\tau}{2}\epsilon_{ijkl}\langle \Bar{k}\Bar{l}\rangle, 
\end{align}
where $E=p_1+p_2+p_3+p_4$ is the total energy. The spinor helicity variables result can thus be obtained by (where the $C$ matrix is gauge-fixed), 
\begin{align}
    \mathbf\Psi_4^{-+-+}=(2\pi)^3\delta^3(\sum_{i=1}^{4}p_i^\mu)\int \frac{d\tau}{2\pi i}\hat{\delta}(C\cdot\Xi^{-+-+})\mathcal{F}_4^{-+-+}(C)|_{c_{ij}=\frac{\langle i j\rangle}{E}+\frac{\tau}{2}\epsilon_{ijkl}\langle \Bar{k}\Bar{l}\rangle}.
\end{align}
This quantity has many poles in the $\tau$ plane. For the flat space limit at hand,  however,  only the $\tau=0$ pole contributes. In the language of minors,  we have $S+T+U=-2 E \tau$ where $E=p_1+p_2+p_3+p_4$ is the total energy and the flat space limit is $E\to 0$. We find, 
\begin{align}\label{psi4flatlimit}
    &\hat A_4^{-+-+}=\lim_{E\to 0}\mathbf\Psi_4^{-+-+}\notag\\&=\frac{\langle 13\rangle^3}{8\langle 12\rangle\langle 23\rangle\langle 3 4\rangle\langle 4 1\rangle}\bigg(\langle 2 4\rangle-\Bar{\xi}_1\xi_2\langle 41\rangle-\Bar{\xi}_1\xi_4\langle 12\rangle-\xi_2\Bar{\xi}_3\langle 3 4\rangle+\Bar{\xi}_3\xi_4\langle 2 3\rangle+\langle 13\rangle \Bar{\xi}_1\xi_2\Bar{\xi}_3\xi_4\bigg).
\end{align}
On the other hand the flat space $\mathcal{N}=1$ SYM on-shell $(-+-+)$ scattering super-amplitude computed in ~\cite{Elvang:2011fx} is given by 
\begin{align}\label{A4flat}
    &\tilde{A}_4^{-+-+}\notag\\&=\frac{\langle 13\rangle^3}{8\langle 12\rangle\langle 23\rangle\langle 3 4\rangle\langle 4 1\rangle}\bigg(\langle 1 3\rangle+\Bar{\eta}_1\eta_2\langle 2 3\rangle-\Bar{\eta}_1\eta_4\langle 3 4\rangle-\eta_2\Bar{\eta}_3\langle 12\rangle-\Bar{\eta}_3\eta_4\langle 4 1\rangle+\Bar{\eta}_1\eta_2\Bar{\eta}_3\eta_4 \langle 24\rangle\bigg).
\end{align}
It is easy to show that expressions in \eqref{psi4flatlimit} and \eqref{A4flat} are identical. This can be shown by making the following change of variables\footnote{Essentially, $\Bar{\eta}$ and $\eta$ are the same as $\xi$ and $\Bar{\xi}$ respectively, as we can see from the definition of our half-Fourier transform \eqref{xitoxibar}.},
\begin{align}
    {\hat A}_4^{-+-+}&=\int d\Bar{\eta}_1e^{\Bar{\eta}_1\Bar{\xi}_1}\int d\eta_2 e^{-\eta_2\xi_2}\int d\Bar{\eta}_3 e^{\Bar{\eta}_3\Bar{\xi}_3}\int d\eta_4 e^{-\eta_4\xi_4} {\tilde A}_4^{-+-+}\notag\\
    &=\frac{\langle 13\rangle^3}{8\langle 12\rangle\langle 23\rangle\langle 3 4\rangle\langle 4 1\rangle}\bigg(\langle 2 4\rangle-\Bar{\xi}_1\xi_2\langle 41\rangle-\Bar{\xi}_1\xi_4\langle 12\rangle-\xi_2\Bar{\xi}_3\langle 3 4\rangle+\Bar{\xi}_3\xi_4\langle 2 3\rangle+\langle 13\rangle \Bar{\xi}_1\xi_2\Bar{\xi}_3\xi_4\bigg).
\end{align}
This gives
\begin{align}
    \lim_{E\to 0}\mathbf\Psi_4^{-+-+}={\tilde A}_4^{-+-+}, 
\end{align}
perfectly matching the flat space $\mathcal{N}=1$ SYM super-amplitude,  establishing the flat space limit in super-space.

Let us also discuss another extremely important fact, which is the R symmetry enhancement from $\mathbb{Z}_2$ to $U(1)$ in the flat limit. The AdS$_4$ result for the correlator is \eqref{psi4gen}. We note that under $\xi_i\to r \xi_i, \Bar{\xi}_i\to \frac{\Bar{\xi}_i}{r}$, there are exactly two discrepant terms viz the ones with the Grassmann structures $\Bar{\xi}_1\Bar{\xi}_3$ and $\xi_2\xi_4$. However, in the flat space limit, it is these two structures which precisely drop out, thereby yielding \eqref{psi4flatlimit} which is invariant under the opposite scaling of the $\xi_i$ and the $\Bar{\xi}_i$. Therefore, the non-trivial $U(1)$ $R-$symmetry group emerges as desired in the flat space limit.

\section{Summary and Discussion}\label{sec:Discussion}
In this paper, we developed a superspace extension of the orthogonal Grassmannian to study $\mathcal{N}=1$ superconformal theories with a focus on making manifest the $\text{OSp}(1|4)$ 3-d super-conformal algebra using the Grassmannian. As a primary application of the super-orthogonal Grassmannian, we demonstrate that the color-ordered non-abelian gluino correlator—constructed purely from $s$- and $t$-channel exchange diagrams—is sufficient to reconstruct the tree-level Yang-Mills four-point function, including contact terms, through the enforcement of supersymmetric constraints. We further establish the universality of these results by showing that the component Yang-Mills correlators coincide with those of non-supersymmetric theories, thereby highlighting the utility of supersymmetric methods for obtaining results with broader applicability. Finally, we validate this superspace formalism by taking the flat-space limit and recovering established scattering amplitudes.
There are some interesting future works that we would like to explore.    


\subsection*{Higher supersymmetry}
The orthogonal Grassmannian OGr(n, 2n) made its first appearance in the study of scattering amplitudes in the three-dimensional $\mathcal{N}$=6 ABJM theory ~\cite{Huang:2014xza,  Huang:2013owa,  Kim:2014hva}. This raises the natural question of whether the OGr(n, 2n) can be systematically extended to theories with higher supersymmetry,  ultimately targeting $\mathcal{N}$=4 higher supersymmetry,  revealing the geometrical structure to 3-d CFT. Building on our previous work on supersymmetry~\cite{Jain:2023idr,  Bala:2025qxr},  we construct the super OGr(n, 2n) for $\mathcal{N}=2, 3, 4$ in our companion paper\cite{Bala:2026bdx}.  Higher supersymmetry should also allow us to obtain graviton AdS amplitudes easily. 

\subsection*{Spinning Correlators for higher spin theories }
Recently, a paper on Vasiliev theory \cite{De:2026shn} appeared where the authors bootstrap a scalar correlator using the orthogonal Grassmannian and obtained very simple results. It would be interesting to see if supersymmetry or the use of higher spin equations or other techniques \cite{Maldacena:2011jn,Li:2019twz,Jain:2022ajd,Jain:2023juk} can be useful in obtaining spinning correlators in these theories, in the Grassmannian framework.

\subsection*{Higher point correlators}
The present work focuses on three- and four-point supercorrelators. We have bootstrapped the supercorrelators using factorization (unitarity) in the bulk, and the same principle should extend to higher-point supercorrelators. On the AdS$_4$ side,  some progress has been made in the computation of higher-point correlators ~\cite{Albayrak:2018tam, Albayrak:2019yve, Jepsen:2019svc} and it would be interesting to pursue this direction.


\subsection*{BCFW recursion}
One of the most powerful tools for computing scattering amplitudes is the BCFW recursion relation,  which expresses n-point amplitudes in terms of products of lower-point amplitudes. There also exist BCFW recursion relations in the (super-)Twistor space approach to flat space scattering amplitudes~\cite{Mason:2009sa, Arkani-Hamed:2009hub}. We have shown that 3d superconformal correlators have the correct flat space limit,  establishing a connection between the CFT correlators and scattering amplitudes. It would be interesting to implement the BCFW in CFT also. Since factorization is the key step in getting BCFW recursion,  and the implementation of factorization is very easy in the Grassmannian space,  it would be interesting to get the BCFW relation in the Grassmannian space.

\subsection*{Higher Dimensional Grassmannian}
    It would be interesting to derive the (super-)Grassmannian construction for higher-dimensional CFTs. There is a lot of interesting literature on the computation/bootstrap of AdS$_5$ correlators see for instance  \cite{Alday:2022lkk,  Alday:2023mvu}. It would be fascinating to make a connection to such constructions.

\section*{\Large Acknowledgment}
AB acknowledges a UGC-JRF fellowship. AAR acknowledges a CSIR-JRF fellowship.\\ D K.S. would like to thank Saurabh Pant for many discussions over the years on (supersymmetric) scattering amplitudes.

\newpage
\appendix

\section{Construction of super-currents in Grassmann twistor variables}\label{app:Grassmanntwistorconstruction}
In this appendix,  we review the construction of $\mathcal{N}=1$  momentum superspace and super-spinor helicity,  Grassmann twistor variables ~\cite{Jain:2023idr}. 
The observables of interest  are correlation functions of conserved super-currents,  which have the following superfield expansion:
\begin{align}\label{momentumspacecurrent}
    \mathbf{J}^{a_1\cdots a_{2s}}_s(p, \theta)=J_s^{a_1\cdots a_{2s}}(p)+\theta_b J_{s+\frac{1}{2}}^{ba_1\cdots a_{2s}}(p)+\frac{\theta^2}{4}p^{a_1}_b J_{s}^{b a_2\cdots a_{2s}}(p).
\end{align}
We trade the momentum vector for a symmetric bi-spinor, 
\begin{align}\label{pmatrix}
    p_{ab}=(\sigma^\mu)_{ab}p^\mu=\lambda_{(a}\Bar{\lambda}_{b)}.
\end{align}
We also express the Grassmann spinor $\theta^a$ in the basis of the spinor helicity variables $(\lambda, \Bar{\lambda})$ as follows:
\begin{align}\label{superSH}
    \theta^a=\frac{\eta \Bar{\lambda}^a+\Bar{\eta}\lambda^a}{2p}.
\end{align}
The two independent components of the super-current \eqref{momentumspacecurrent} can be obtained by contracting with the polarization spinors as follows:
\begin{align}\label{Jspm}
    \mathbf{J}_s^{\pm}(\lambda, \Bar{\lambda}, \theta)=\zeta_{\pm a_1}\cdots \zeta_{\pm a_{2s}}\mathbf{J}_s^{a_1\cdots a_{2s}}(p, \theta), ~\zeta_{+a}=\frac{\lambda_a}{\sqrt{p}}, \zeta_{-a}=\frac{\Bar{\lambda}_a}{\sqrt{p}}.
\end{align}
Using the super-spinor helicity formula \eqref{superSH} we find, 
\begin{align}
    &\mathbf{J}_s^{-}(\lambda, \Bar{\lambda}, \eta, \Bar{\eta})=e^{-\frac{\eta\Bar{\eta}}{4}}J_s^{-}(\lambda, \Bar{\lambda})+\frac{\Bar{\eta}}{2\sqrt{p}}J_{s+\frac{1}{2}}^{-}(\lambda, \Bar{\lambda}), \notag\\
    &\mathbf{J}_s^{+}(\lambda, \Bar{\lambda}, \eta, \Bar{\eta})=e^{\frac{+\eta\Bar{\eta}}{4}}J_s^{+}(\lambda, \Bar{\lambda})+\frac{\eta}{2\sqrt{p}}J_{s+\frac{1}{2}}^{+}(\lambda, \Bar{\lambda}).
\end{align}
The structure of the exponential motivates one to perform a Grassmann half-Fourier transform with respect to $\eta$ or $\Bar{\eta}$. We construct for negative helicity, 
\begin{align}
\mathbf{J}_s^{-}(\lambda, \Bar{\lambda}, \Bar{\xi})&=4 \int d(\Bar{\chi}-\Bar{\eta})\int d\eta e^{-\frac{\Bar{\chi}\eta}{4}}\mathbf{J}_s^{-}(\lambda, \Bar{\lambda}, \eta, \Bar{\eta})\notag\\
&=J_s^{-}(\lambda, \Bar{\lambda})+\frac{\Bar{\xi}}{\sqrt{2 p}}J_{s+\frac{1}{2}}^{-}(\lambda, \Bar{\lambda}).
\end{align}
For positive helicity on the other hand we consider, 
\begin{align}
\mathbf{J}_s^{+}(\lambda, \Bar{\lambda}, \xi)&=4\int d(\chi-\eta)\int d\Bar{\eta}e^{-\frac{\chi\Bar{\eta}}{4}}\mathbf{J}_s^{+}(\lambda, \Bar{\lambda}, \eta, \Bar{\eta})\notag\\
&=J_s^{+}(\lambda, \Bar{\lambda})+\frac{\xi}{\sqrt{2p}}J_{s+\frac{1}{2}}^{+}(\lambda, \Bar{\lambda}).
\end{align}
where $\Bar{\xi}=\frac{\Bar{\chi}-\Bar{\eta}}{2\sqrt{2}}$ and $\xi=\frac{\chi-\eta}{2\sqrt{2}}$.

\section{The Geometry of the Orthogonal Grassmannian}\label{app:GeometryOfGrassmannian}
Before discussing the geometry of the Orthogonal Grassmannian,  let us fix the notations and conventions first.\\
Given the little group scaling of $\{\lambda_1, \cdots, \lambda_n, \Bar{\lambda}_1, \cdots\Bar{\lambda}_n\}\to \{\rho_1\lambda_1, \cdots, \rho_n\lambda_n, \frac{\Bar{\lambda}_1}{\rho_1}, \cdots, \frac{\Bar{\lambda}_n}{\rho_n}\}$,  for the expression \(C\cdot \Lambda\) to be little group invariant,  the \(C\) matrix should transform as  $\{c_1, \cdots, c_n, {c}_{n+1}, ~\cdots {c}_{2n}\}\to \{\frac{c_1}{\rho_1}, \cdots, \frac{c_n}{\rho_n},  ~\rho_1 {c}_{n+1}, \cdots, \rho_n {c}_{2n}\}$. Therefore,  we label the \(C\) matrix as 
 \begin{equation}
     \{c_1, \cdots, c_n, {c}_{n+1}, ~\cdots {c}_{2n}\} \equiv \{ \bar 1,  \cdots \bar n,  1, \cdots n \},
 \end{equation}
 and the notation \((i_1 i_2\cdots i_n)\) where \(i\) can be either barred or unbarred,  denotes
 \begin{equation}
     (i_1 i_2\cdots i_n) \equiv \det(\{i_1,  i_2, \cdots i_n  \}).
 \end{equation}
 As an example,  in the case of four-point functions,  \((12\bar 3 \bar 4) = \det(\{1,  2, \bar 3,  \bar 4\}) = \det(\{c_5,  c_6,  c_3,  c_4\})\).\\
With the notations and conventions in place,  let us proceed with the discussion.\\
The solution space to the constraint 
\begin{equation}
    C\cdot Q\cdot C^T = 0 \implies (\bar 1 \bar 2 \bar 3 \cdots \bar n)(1 \bar 2 \bar 3\cdots \bar n) = 0,
\end{equation}
has two disconnected branches,  one with \( (\bar 1 \bar 2 \bar 3 \cdots \bar n) = 0\) and the other with \((1 \bar 2 \bar 3\cdots \bar n) = 0\) which are called left and right branches respectively. \\
For 2-point correlators,  a convenient gauge fixing on the right branch is the one used in \eqref{2pointRightBranch}
\begin{equation}
    C=\begin{pmatrix}
        1&0&0&c_{12}\\
        0&1&-c_{12}&0
    \end{pmatrix}.
\end{equation}
while on the left branch a gauge fixed \(C\) matrix will be of the form 
\begin{equation}
    C=\begin{pmatrix}
        1&c_{12}&0&0\\
        0&0&-c_{12}&1
    \end{pmatrix}.
\end{equation}
For 3-point correlators,  conveniently,  the left and right branches coincide with the gauge-fixing choice of setting the last \(3\) columns to identity and the first \(3\) columns to identity,  respectively. Therefore,  on the right branch,  a gauge-fixed matrix would look like \eqref{3pointRightBranch}
        \begin{align}
    C=\begin{pmatrix}
        1&0&0&0&-c_{12}&-c_{13}\\
        0&1&0&c_{12}&0&-c_{23}\\
        0&0&1&c_{13}&c_{23}&0
    \end{pmatrix}.
\end{align}
While on the left branch, it would look like \eqref{3pointLeftBranch}
 \begin{align}
    C=\begin{pmatrix}
        0&-c_{12}&-c_{13}&1&0&0\\
        c_{12}&0&-c_{23}&0&1&0\\
        c_{13}&c_{23}&0&0&0&1
    \end{pmatrix}.
\end{align}
This is not special only for \(3\) points,  but is true for general odd \(n\),  since any antisymmetric \(n\times n\) matrix has 0 determinant,  and therefore,  setting last \(n\) columns to identity,  which forces the first \(n\) columns to be an antisymmetric matrix,  will automatically put the \(C\) matrix on the left branch,  and vice versa.\\
For SYM theory with no higher-order interactions,  all four-point functions with an even number of + helicities are supported only on the right branch,  while the correlators with an odd number of helicities are supported only on the left branch. In this paper,  we primarily deal with four-point correlators supported on the right branch,  and we have used the gauge fixing as given in \eqref{rightBranchGaugeFixing}
\begin{align}
    C=\begin{pmatrix}
        1&0&0&0&0&-c_{12}&-c_{13}&-c_{14}\\
        0&1&0&0&c_{12}&0&-c_{23}&-c_{24}\\
        0&0&1&0&c_{13}&c_{23}&0&-c_{34}\\
        0&0&0&1&c_{14}&c_{24}&c_{34}&0
    \end{pmatrix}.
\end{align}
If one were to calculate,  say,  the 4-point \((+++-)\) correlator,  one would have to work in the left branch,  and an example of a gauge-fixed solution that lies in the left branch is as follows 
\begin{equation}
    C = \begin{pmatrix}
         1 & 0 & 0 & -c_{14} & 0 & -c_{12} & -c_{13} & 0 \\
         0 & 1 & 0 & -c_{24} & c_{12} & 0 & -c_{23} & 0 \\
         0 & 0 & 1 & -c_{34} & c_{13} & c_{23} & 0 & 0 \\
         0 & 0 & 0 & 0 & c_{14} & c_{24} & c_{34} & 1 \\
    \end{pmatrix}.
\end{equation}
\\
See that if we write \(C = (C_L~ | ~C_R)\),  then the constraint \(C\cdot Q\cdot C^T = 0\) imposes 
\begin{equation}
    C_L C_R^T + C_R C_L^T = 0 \implies C_L C_R^T = A,
\end{equation}
where \(A\) is an antisymmetric matrix. On the right branch,  since \((\bar 1\bar 2\cdots \bar n) = \det C_L \ne 0\),  one can simply choose some \(A\) and solve for \(C_R\) as \(C_R = -A (C_L^{-1})^T\). Since \(A\) is also an invertible matrix (in the case of even \(n\)),  \(C_R\) also has a non-zero determinant. \\
Therefore,  a general solution on the right branch (up to \(\mathbb{GL}(n)\) transformations) will be of the form
\begin{equation}
    C = \begin{pmatrix}
        C_L &  -A (C_L^{-1})^T.
    \end{pmatrix}.
\end{equation}
Using this general solution,  one can show that the following relations are true on the right branch
\begin{equation}
    \begin{split}
        & (\bar 3\bar 4 3 4) = (\bar 1\bar 212) =S,\quad
        (\bar 2\bar 3 23) = (\bar 1\bar 4 14) = T,\quad
         (\bar 2\bar 4 24) = (\bar 1\bar 3 13) = U.
    \end{split}
\end{equation}
On the left branch,  we have \((\bar 1\bar 2 \cdots \bar n) = \det C_L = 0\) which implies that \(\mathrm{Rank}(C_L) = k<n\). Therefore,  we can always go to a basis where 
\begin{equation}
    C_L = \begin{pmatrix} M_{k\times k} & (M N)_{k\times (n-k) } \\ 0_{(n-k)\times k} & 0_{(n-k) \times (n-k)} \end{pmatrix},
\end{equation}
where \(M\) is some \(k\times k\) matrix with non-zero determinant,  and \(N\) is a \(k \times (n-k)\) matrix.\\
Then,  the constraint \(C_L C_R^T  = -C_R C_L^T\) translates to the requirement that
\begin{equation}
    \begin{pmatrix} M_{k\times k} & (M N)_{k\times (n-k) } \\ 0_{(n-k)\times k} & 0_{(n-k) \times (n-k)} \end{pmatrix} \begin{pmatrix} (C_{R}^{(11)})^T_{k\times k} & (C_R^{(21)})^T_{k\times (n-k)} \\ (C_R^{(12)})^T_{(n-k)\times k} & (C_R^{(22)})^T_{(n-k)\times (n-k)} \end{pmatrix},
\end{equation}
must evaluate to an antisymmetric matrix. \\
Expanding the product gives:
\begin{equation}
    \begin{pmatrix} M (C_R^{(11)})^T + M N (C_R^{(12)})^T & M (C_R^{(21)})^T + M N (C_R^{(22)})^T \\ 0 & 0 \end{pmatrix}.
\end{equation}
For this to be antisymmetric,  the top-right block must be identically zero. Since \(M\) is invertible,  this imposes \(C_R^{(21)} = -C_R^{(22)} N^T\). Furthermore,  the top-left block must equal some antisymmetric matrix \(A_{k \times k}\),  which yields \(C_R^{(11)} = -A (M^{-1})^T - C_R^{(12)} N^T\). \\
Therefore,  a general solution on the left branch is given by 
\begin{equation}
    \begin{pmatrix} M_{k\times k} & (M N)_{k\times n-k} & \left(-A (M^{-1})^T - C_R^{(12)} N^T\right)_{k\times k} & (C_R^{(12)})_{k\times n-k} \\ 0_{n-k\times k} & 0_{n-k \times n-k} & \left(-C_R^{(22)} M^T\right)_{n-k\times k} & (C_{R}^{(22)})_{n-k\times n-k} \end{pmatrix}.
\end{equation}
Using this general solution,  one can show that on the left branch,   
\begin{equation}
    \begin{split}
        & (\bar 3\bar 4 3 4) = -(\bar 1\bar 212) =-S,\quad
        (\bar 2\bar 3 23) = -(\bar 1\bar 4 14) = -T,\quad
        (\bar 2\bar 4 24) = -(\bar 1\bar 3 13) = -U.
    \end{split}
\end{equation}

\section{The Grassmann Operator Delta function}\label{app:NewDeltaFunction}
In this appendix,  we discuss the operator-valued Grassmann delta function that appears in \eqref{deltaCdotX}. We begin with the ordinary Grassmann Dirac delta function and generalize it to suit our construction. 

For a single Grassmann variable $\xi$ we have, 
\begin{align}
    \delta(\xi)=\xi.
\end{align}
This quantity satisfies 
\begin{align}
    \int d\xi \delta(\xi)=1, ~~\delta(a \xi)=a \delta(\xi) , ~~\xi\delta(\xi)=0, 
\end{align}
where we used the fact that Grassmann variables square to zero.
It can be expressed using Schwinger parameterization as follows:
\begin{align}
    \delta(\xi)=\int d\theta e^{\theta\xi}, 
\end{align}
as can be checked using the above properties. The generalization to an $n$ component Grassmann vector $\xi^{i}$ is natural in this language and takes the form, 
\begin{align}
    \delta^n(\xi)=\int d^n \theta e^{\theta_i \xi^i}=\frac{1}{n!}\epsilon^{i_1\cdots i_n}\xi_{i_1}\cdots \xi_{i_n}=\xi_1\cdots \xi_n, 
\end{align}
where $d^n\theta=d\theta_n\cdots d\theta_1$ and we used  the Grassmann variable anti-commutation relation $\{\xi_i, \xi_j\}=0$. This quantity satisfies, 
\begin{align}
    \int d^n \xi~\delta^n(\xi)=1.
\end{align}
Let's now consider the operator-valued delta function that forms the core of our formalism,  viz. $\delta(C\cdot\Xi^{h_1\cdots h_n})$. Before we proceed,  it will be useful to define, 
\begin{align}
    \eta_i=C_{iA}\Xi^A.
\end{align}
The $\eta_i$ variables satisfy, 
\begin{align}
    \{\eta_i, \eta_j\}=C_{iA}C_{jB}\{ \Xi^{A, h_1, \cdots h_n}, \Xi^{B, h_1\cdots h_n}\}=C_{iA}C_{jB}Q^{AB}.
\end{align}
In general,  the RHS of the above equation is non-zero. However,  in the context of the orthogonal Grassmannian,  it is zero as a consequence of the delta function constraints enforcing $C.Q.C^T=0$. Therefore, 
\begin{align}
    \{\eta_i, \eta_j\}=0~~\text{when}~~C.Q.C^T=0.
\end{align}
In other words,  the $\eta_i$,  which are operators,  behave like ordinary anti-commuting Grassmann quantities at the support of the orthogonality constraint. Therefore,  we have, 
\begin{align}
    \delta((C\cdot\Xi)_i)=\delta(\eta_i)=\int d^n \theta~e^{\theta^i\eta_i}=\frac{\epsilon^{i_1\cdots i_n}}{n!}\eta_{i_1}\cdots \eta_{i_n}=\eta_1\cdots\eta_n.
\end{align}
As a consequence of the anti-commutation relation,  we have inside the orthogonal Grassmannian integrand  the following identity:
\begin{align}
    \eta_k\delta(\eta_i)=0, ~k\in\{1, \cdots, n\}.
\end{align}
This property is essential to ensure the supersymmetry and special super-conformal Ward identities of the super-orthogonal Grassmannian.

\section{The Bi-linear Grassmann  Exponential}\label{app:bilinearUblock}
    In this appendix, we show that there exists another solution to the super-conformal generators $\mathcal{Q} ~\text{and}~ \mathcal{S}$. \\    
    In sub-section (\ref{subsec:susyfromgenerator}), we showed that 
    \begin{align}
\boxed{\mathbf\Phi_n^{h_1\cdots h_n}(C\cdot\Lambda,  C\cdot\Xi^{h_1\cdots h_n},  C)= \hat\delta(C\cdot\Xi^{h_1\cdots h_n}) \delta(C\cdot \Lambda)\delta(C\cdot Q\cdot C^T)\mathcal{F}(C)1} 
\end{align}
 solves the superconformal generators. \\
 Now, let us consider another ansatz for the super correlator:
  \begin{align}
\boxed{
\begin{aligned}
\tilde{\mathbf\Phi}_n^{h_1\cdots h_n}(C\cdot\Lambda,  \Xi^{h_1\cdots h_n},  C)=\int\bigg( \prod_{i_k} d\chi_{i_k}~ e^{-\chi_{i_k} \bar\chi_{i_k}} \bigg) \hat\delta(C\cdot\Xi^{h_1,\cdots, -h_{i_1},-h_{i_2},\cdots, h_n})~~ \\\delta(C\cdot \Lambda)\delta(C\cdot Q\cdot C^T)\mathcal{F}(C)1~~,
\end{aligned}}
\end{align}
where $i_k$ are the legs we choose to Fourier transform, and the notation $\Xi^{h_1,\cdots, -h_{i_1}, -h_{i_2},\cdots, h_n}$ indicates that the legs which are to be Fourier transformed are taken to have the opposite helicity. The $\chi$ and its Fourier conjugate $\bar\chi$ represent $\xi$ or $\bar\xi$ depending on the Grassmann ``phase'' vector $\Xi^{h_1,\cdots -h_{i_1}, -h_{i_2},\cdots , h_n}$ as given in \eqref{XiVectorNeq1}.

The action of $\mathcal{S}$ as given in \eqref{Saction1}, on the above super correlator $\tilde{\mathbf\Phi}_n^{h_1\cdots h_n}$ is given by
      \begin{align}
          \mathcal{S}(\Lambda,&\Xi^{h_1\cdots h_n}) \tilde{\mathbf\Phi}_n^{h_1\cdots h_n} \notag \\ 
          &= \mathcal{S}(\Lambda,\Xi^{h_1\cdots h_n}(\bar\chi)) \int \prod_{i_k}d\chi_{i_k}~ e^{-\chi_{i_k} \bar\chi_{i_k}}  \hat\delta(C\cdot\Xi^{h_1,\cdots,-h_i,\cdots, h_n}) \delta(C\cdot \Lambda)\delta(C\cdot Q\cdot C^T)\mathcal{F}(C)1 \notag\\
          & =  \int \prod_{i_k} d\chi_{i_k}~ \mathcal{S}(\Lambda,\Xi^{h_1\cdots h_n}(\bar\chi))\big(e^{-\chi_{i_k} \bar\chi_{i_k}}  \hat\delta(C\cdot\Xi^{h_1,\cdots,-h_{i_1},-h_{i_2}, \cdots, h_n}) \delta(C\cdot \Lambda)\delta(C\cdot Q\cdot C^T)\mathcal{F}(C)1\big)\notag \\
          & =  \int \prod_{i_k}d\chi_{i_k}~ e^{-\chi_{i_k} \bar\chi_{i_k}} \mathcal{S}(a\Lambda,\Xi^{h_1,\cdots,-h_{i_1},-h_{i_2},\cdots, h_n}(\chi))\big( \hat\delta(C\cdot\Xi^{h_1,\cdots,-h_{i_1}, -h_{i_2},\cdots, h_n})\notag\\&\qquad\qquad\qquad\qquad\qquad\qquad\qquad\qquad\qquad\qquad\qquad\qquad\quad \delta(C\cdot \Lambda)\delta(C\cdot Q\cdot C^T)\mathcal{F}(C)1\big) \notag\\
          &=  \int \prod_{i_k} d\chi_{i_k}~ e^{-\chi_{i_k} \bar\chi_{i_k}} \mathcal{S}(a\Lambda,\Xi^{h_1,\cdots,-h_{i_1},-h_{i_2},\cdots, h_n}(\chi))\big(\mathbf\Phi_n^{h_1,\cdots,-h_{i_1},-h_{i_2},\cdots, h_n}\big) \notag\\
          &=0~~,
      \end{align}
where going from $2^{nd}$ line to the $3^{rd}$ line, the $\mathcal{S}$ generator got Fourier transformed and the action of it on $\mathbf\Phi_n^{h_1,\cdots, -h_{i_1}, -h_{i_2},\cdots, h_n}$ is zero as shown in the subsection (\ref{subsec:susyfromgenerator}). 

We have further observed empirically that the Grassmann parity of the $\hat{\mathcal U}(C,\Xi)$ blocks is determined by the parity of the number of Grassmann Fourier transforms. In particular, if $\hat\delta(C\cdot \Xi)$ is Grassmann-even, then Fourier transforming an even number of Grassmann variables produces a Grassmann-even block proportional to $\hat\delta(C\cdot \Xi)$, whereas Fourier transforming an odd number of variables produces Grassmann-odd blocks, which are all proportional to one another. Similarly, if $\hat\delta(C\cdot \Xi)$ is Grassmann-odd, an odd number of Fourier transforms produces a Grassmann-even block, while an even number produces a  Grassmann-odd block proportional to $\hat\delta(C\cdot \Xi)$. This proportionality follows from the uniqueness of the solution to the supersymmetric Ward identities \(\mathcal Q\) and \(\mathcal S\). 

Therefore, the full solution to the supercorrelator is:
\begin{align}
    \mathbf{\Psi}_n^{h_1\cdots h_n}=\int \frac{d^{n\times 2n}C}{\text{Vol}(\mathbb{GL}(n))}\delta(C\cdot Q\cdot C^T)\delta(C\cdot\Lambda)\Bigg[\hat{\delta}(C\cdot{{\Xi}}^{h_1\cdots h_n}) \mathcal{F}^{h_1\cdots h_n}(C)\qquad\qquad \notag\\+\hat{\mathcal{U}}(C,{{\Xi}}^{h_1\cdots h_n}) \mathcal{G}^{h_1\cdots h_n}(C)\Bigg], 
\end{align}
where, \begin{align}\label{Ublock}
    \hat{\mathcal{U}}^{h_1\cdots h_n}(C,{\Xi}^{h_1\cdots h_n}) =\int \bigg(\prod_{i_k}d^\mathcal{N}{\chi}_{i_k}~e^{-\chi_{i_k}\cdot\,\Bar{\chi}_{i_k}}\bigg)~\hat{\delta}(C\cdot{{\Xi}}^{h_1,\cdots, -h_{i_1}, -h_{i_2},\cdots, h_n}).
\end{align}

One can note in the above expression that the two blocks $\hat\delta(C\cdot\Xi)$ and $\hat{\mathcal{U}}(C,\Xi)$ independently solve the superconformal generators; then one can ask how to choose one block over the other. It is the choice of super-current expansion that will enforce a choice of one of the blocks, with the other one not contributing. 

In the $\mathcal{N}=1$ half-integer multiplet we consider in this paper, our variable choice selects the superconformal building block $\hat\delta(C\cdot\Xi)$ throughout the paper rather than $\hat{\mathcal{U}}$. The correlators that fall in the \(\hat{\mathcal{U}}\) block are kinematically not allowed, and therefore, the entire block is irrelevant for our discussions.
  The $\hat{\mathcal{U}}$ block will be of importance in higher $\mathcal{N}=2,3,4$ supersymmetry, which will be discussed in detail in the companion paper \cite{Bala:2026bdx}.

\section{Extension to integer super-currents}\label{app:Neq1Integerspins}
In this appendix we show how we can extend the formalism we have developed for half-integer spin super-currents to integer spin super-currents.\\ 
The integer spin super-current is given as follows
\begin{align}
    \mathbf{J}_s^{+}(\lambda, \Bar{\lambda}, \Bar{\xi})=\Bar{\xi}J_s^{+}(\lambda, \Bar{\lambda})+\frac{1}{\sqrt{2}}J_{s+\frac{1}{2}}^{+}(\lambda, \Bar{\lambda}), ~~\mathbf{J}_s^{-}(\lambda, \Bar{\lambda}, \xi)=\xi J_s^{-}(\lambda, \Bar{\lambda})+\frac{1}{\sqrt{2}}J_{s+\frac{1}{2}}^{-}(\lambda, \Bar{\lambda}), ~s\in \mathbb{Z}_{>0}, 
\end{align}
 For this,  we require 
\begin{align}
    \tilde{\Xi}^{h_1\cdots h_n}=\Xi^{-h_1\cdots -h_n}.
\end{align}
which is simply due to our choice of variables.\\
After the constraint from the super-conformal generators,  the integer multiplet supercorrelator takes the same form as \eqref{SuperGrassmannianNeq1} with the Grassmann phase vector as $\tilde\Xi^{h_1\cdots h_n}$,  i.e.:
 \begin{align}\label{SuperGrassmannianNeq1integer}
   \boxed{\mathbf\Psi_n^{h_1\cdots h_n}=\int \frac{d^{n\times 2n}C}{\text{Vol}(\mathbb{GL}(n))}\delta(C\cdot Q\cdot C^T)\delta^{n}(C\cdot\Lambda)\hat\delta^{n}(C\cdot\tilde\Xi^{h_1\cdots h_n})\mathcal{F}^{h_1\cdots h_n}(C).}
   \end{align}
Similarly,  one can write the ansatz for $\mathcal{F}(C)$ and get all the supercorrelators in the integer multiplet following the procedure in the main text. For the current choice of the super-multiplet the other solution discussed in appendix \ref{app:bilinearUblock} is trivial since it multiplies a component correlator that is zero. 

\section{Extension to super-scalars}\label{app:scalar}
In this brief appendix, we present the extension of our formalism to super-scalar operators. The super-scalar multiplet contains two scalars $O_1,O_2$ and a spin-$\frac{1}{2}$ operator $O_{\frac{1}{2}}^{a}$. 
\begin{align}
    \mathbf{J}_0=O_1+\theta_a O_{\frac{1}{2}}^a+\frac{\theta^2}{4}O_2.
\end{align}
Like in Appendix \ref{app:Grassmanntwistorconstruction}, we convert this quantity to super-spinor helicity variables, obtaining
\begin{align}
    \mathbf{J}_0(\lambda,\Bar{\lambda},\eta,\Bar{\eta})=O_1+\frac{\eta}{2}O_{\frac{1}{2}}^{+}+\frac{\Bar{\eta}}{2}O_{\frac{1}{2}}^{-}+\frac{\eta\Bar{\eta}}{4}O_2,
\end{align}
where all operators are the rescaled ones, such that the SCT operator has a simple action. We now construct two superfields in the Grassmann twistor variables exactly following the spinning construction in appendix \ref{app:Grassmanntwistorconstruction}. This results in,
\begin{align}
    &\mathbf{J}_{0}^{+}=O+\frac{\xi}{\sqrt{2}}O_{\frac{1}{2}}^{+}\notag\\
    &\mathbf{J}_0^{-}=\Bar{O}+\frac{\Bar{\xi}}{\sqrt{2}}O_{\frac{1}{2}},
\end{align}
However, it is more convenient to work with the opposite Grassmann twistor convention viz,
\begin{align}
    &\mathbf{J}_{0}^{+}=\Bar{\xi}O+\frac{1}{\sqrt{2}}O_{\frac{1}{2}}^{+}\notag\\
    &\mathbf{J}_0^{-}=\xi\Bar{O}+\frac{1}{\sqrt{2}}O_{\frac{1}{2}},
\end{align}
The scalars are given by,
\begin{align}
    O=O_1+O_2,~\Bar{O}=O_1-O_2.
\end{align}
Since $O_2$ is odd under parity, we find that $O$ and $\Bar{O}$ transform into one another under a $CPT$ transformation. It is now a simple matter to use these superfields to compute correlation functions of these operators, following our general methods for half-integer spin super-currents, setting $s=0$.

\section{Supersymmetry in spinor helicity variables}\label{app:SUSYinSpinorHelicity}
Here we discuss the constraints imposed by supersymmetry on the component correlators in spinor-helicity variables,  and demonstrate the advantage of Grassmannian formalism wherein the differential constraints in spinor-helicity variables become algebraic constraints in Grassmannian.\\
Using the superfield expansion \eqref{Neq1SYMAdSsuperfield},  we can write the 4-point \(\mathbf{J}_{\frac{1}{2}}\) supercorrelator in \((-+-+)\) helicity as 
\begin{equation}\label{SuperCorrelatorExpansion}
\begin{split}
    \langle \mathbf{J}^{-}_{\frac{1}{2}}\mathbf{J}^{+}_{\frac{1}{2}}\mathbf{J}^{-}_{\frac{1}{2}} \mathbf{J}^{+}_{\frac{1}{2}} \rangle = \frac{1}{4} \langle J^- J^+ J^- J^+ \rangle  \bar{\xi }_1 \xi _2  \bar{\xi }_3 \xi _4  + \frac{1}{2} \langle J^- J^+ O_{\frac{1}{2}}^- O_{\frac{1}{2}}^+ \rangle  \bar{\xi }_1 \xi _2 -\frac{1}{2} \langle J^- O_{\frac{1}{2}}^+ J^- O_{\frac{1}{2}}^+ \rangle \bar{\xi }_1 \bar{\xi }_3\\
    ~~~~~~~~+\frac{1}{2}\langle J^- O_{\frac{1}{2}}^+  O_{\frac{1}{2}}^- J^+ \rangle   \bar{\xi }_1 \xi _4 +\frac{1}{2} \langle O_{\frac{1}{2}}^- J^+ J^-  O_{\frac{1}{2}}^+ \rangle  \xi _2 \bar{\xi }_3 + \frac{1}{2} \langle O_{\frac{1}{2}}^- O_{\frac{1}{2}}^+ J^-  J^+ \rangle  \bar{\xi }_3\xi _4  \\
    ~~~~~~~-\frac{1}{2} \langle O_{\frac{1}{2}}^- J^+ O_{\frac{1}{2}}^-  J^+ \rangle  \xi _2 \xi _4+  \langle O_{\frac{1}{2}}^- O_{\frac{1}{2}}^+ O_{\frac{1}{2}}^-  O_{\frac{1}{2}}^+ \rangle. 
\end{split}
\end{equation}
Since the supercorrelator is annihilated by \(\mathcal{Q}\),  we get a constraint of the form
\begin{equation}
   \bigg( \sum_{i=2, 4}\bigg(\Bar{\lambda}_{ia} \xi_i+\lambda_{ia}\frac{\partial}{\partial \xi_i}\bigg) + \sum_{j=1, 3}\bigg(\Bar{\lambda}_{ja}\frac{\partial}{\partial\Bar{\xi}_j}+\lambda_{ja} \Bar{\xi}_j\bigg)\bigg)\langle \mathbf{J}^{-}_{\frac{1}{2}}\mathbf{J}^{+}_{\frac{1}{2}}\mathbf{J}^{-}_{\frac{1}{2}} \mathbf{J}^{+}_{\frac{1}{2}} \rangle = 0.
\end{equation}
This constraint will relate the component correlators,  and we get an expression for the supercorrelator completely in terms of the top and bottom components
\begin{equation}
      \langle \mathbf{J}^{-}_{\frac{1}{2}}\mathbf{J}^{+}_{\frac{1}{2}}\mathbf{J}^{-}_{\frac{1}{2}} \mathbf{J}^{+}_{\frac{1}{2}} \rangle =  \Gamma_{\frac{1}{2}}^{-+-+} ~  \langle O_{\frac{1}{2}}^- O_{\frac{1}{2}}^+ O_{\frac{1}{2}}^-  O_{\frac{1}{2}}^+ \rangle  +  \Gamma_{1}^{-+-+}  ~ \langle J^- J^+ J^- J^+ \rangle ,
\end{equation}
where 
\begin{equation}
   \Gamma_{\frac{1}{2}}^{-+-+}  = 1 + \frac{1}{p_1-p_2+p_3-p_4} \left( \bar{\xi }_1 \xi _2 \left\langle 1 \bar{2}\right\rangle +\bar{\xi }_1 \bar{\xi }_3 \langle 1 3\rangle + \bar{\xi }_1\xi _4 \left\langle 1 \bar{4}\right\rangle + \xi _2 \bar{\xi }_3 \left\langle \bar{2} 3\right\rangle  + \xi _2 \xi _4 \left\langle \bar{2} \bar{4}\right\rangle 
   + \bar{\xi }_3 \xi _4 \left\langle 3 \bar{4}\right\rangle \right)   ,
\end{equation}
and
\begin{equation}
\begin{split}
    \Gamma_{1}^{-+-+} = \frac{1}{4} \bigg(\frac{1}{p_1-p_2+p_3-p_4}(\bar{\xi }_1\xi _2 \left\langle\bar{3} 4\right\rangle -\bar{\xi }_1 \bar{\xi }_3 \langle 24\rangle &+\bar{\xi }_1\xi _4  \left\langle 2   \bar{3}\right\rangle  +\xi _2 \bar{\xi }_3 \left\langle \bar{1} 4\right\rangle \\  &- \xi _2 \xi _4 \left\langle \bar{1} \bar{3}\right\rangle  + \bar{\xi }_3\xi _4 \left\langle \bar{1} 2\right\rangle ) + \bar{\xi }_1 \xi _2 \bar{\xi }_3 \xi _4  \bigg),
\end{split}
\end{equation}
with \(\mathcal{Q}\) annihilating both \(\Gamma_1 \) and \(\Gamma_\frac{1}{2}\) independently. 

The SCT generator \(\mathcal{K}_{ab}\) imposes further constraints,  relating the top and bottom components to each other. However,  it is a second-order differential operator,  and even in the simplest case where \(\mathcal{K}_{ab}\) annihilates the supercorrelator,  one is faced with the challenge of solving a complicated differential equation
\begin{equation}
   \sum_{i=1}^4 \frac{\partial^2}{\partial \lambda_i^{(a}\partial \bar\lambda^{b)}_{i}} \left( \Gamma_{\frac{1}{2}}^{-+-+} ~  \langle O_{\frac{1}{2}}^- O_{\frac{1}{2}}^+ O_{\frac{1}{2}}^-  O_{\frac{1}{2}}^+ \rangle  +  \Gamma_{1}^{-+-+}  ~ \langle J^- J^+ J^- J^+ \rangle   \right) = 0.
\end{equation}
On expanding this, we get six non-trivial linear differential equations for 6 different component correlators. \\
For example, the \( \langle J^- J^+ O_{\frac{1}{2}}^- O_{\frac{1}{2}}^+ \rangle \) correlator which can be read off as the coefficient of \(\bar\xi_1 \xi_2\) in \eqref{SuperCorrelatorExpansion} can be written as  as
\begin{equation}\label{JJOhalOhalfFromtopandbottom}
     \langle J^- J^+ O_{\frac{1}{2}}^- O_{\frac{1}{2}}^+ \rangle  =  \frac{2}{p_1-p_2+p_3-p_4}  \left(\left\langle 1 \bar{2}\right\rangle \langle O_{\frac{1}{2}}^- O_{\frac{1}{2}}^+ O_{\frac{1}{2}}^-  O_{\frac{1}{2}}^+ \rangle + \frac{\left\langle \bar{3}4 \right\rangle}{4} \langle J^- J^+ J^- J^+ \rangle  \right).
\end{equation}
Here the LHS is annihilated by \(\mathcal{K}\), giving a linear differential constraint relating \(\langle J^- J^+ J^- J^+ \rangle  \) and \(\langle O_{\frac{1}{2}}^- O_{\frac{1}{2}}^+ O_{\frac{1}{2}}^-  O_{\frac{1}{2}}^+ \rangle \). Similarly, we can obtain the other five differential constraints.\\
In the Grassmannian formalism, both the above differential constraints are trivially satisfied,  and the entire supercorrelator is fixed completely once the form of \(\mathcal{F}(C)\) is obtained.

\section{\(\mathcal{N} = 1\) super-correlators in other helicities}\label{app:NkMHVSYM}
In this appendix,  we present the expressions for the supercorrelators in other helicities.\\
\textit{Two-point functions—} \\
For two point function,  the results in \((--)\) helicity are as follows
\begin{equation}
    \hat\delta(C\cdot \Xi^{--}) = (\bar 1 1) + (\bar 2 2) + 2\Bar{\xi}_1\Bar{\xi}_2 (12),  ~~    \mathcal{F}^{--}(C)=\frac{({1}{2})^{2s}}{(\Bar{1}1)^{2s}}.
\end{equation}
\textit{Three-point functions —} \\ 
In the main text,  we have worked out the expressions for \((+++)\) and \((++-)\) helicities. Here we present results for \((+--)\) and \((---)\) helicities.\\
For \((+--)\) helicity,  we have 
\begin{align}
    \hat\delta (C\cdot \Xi^{+--}) &=  \frac{1}{2}((\bar{1}{2}\bar 2) - (\bar{1}\bar{3}3))\xi_1 - \frac{1}{2}((\bar{1}21) + (2\bar{3}3))\bar\xi_2 \notag \\
    &\quad + \frac{1}{2}((\bar{1}13) + ({2}\bar23))\bar{\xi}_3 - (\bar{1}{2}3)\xi_1\bar\xi_2\bar{\xi}_3,
\end{align}
and 
 \begin{align}
         { \mathcal{F}^{+--}(C) = \bigg(\frac{(\bar1 23)^{2s_1}(\bar1 21)^{2s_2-2s_1}(\bar131)^{2s_3-2s_1} } {((1\bar1\bar2)(23\bar3))^{s_2+s_3-s_1+\frac{1}{2}}}\bigg)} .
      \end{align}
For \((---)\) helicity,  we have 
\begin{align}
     \hat\delta (C\cdot \Xi^{---}) &=  \frac{1}{2}(({1}{2}\bar 2) - ({1}\bar{3}3))\bar\xi_1 - \frac{1}{2}(({1}2\bar 1) + (2\bar{3}3))\bar\xi_2 \notag \\
    &\quad + \frac{1}{2}(({1}\bar1 3) + ({2}\bar23))\bar{\xi}_3 - ({1}{2}3)\bar\xi_1\bar\xi_2\bar{\xi}_3,
\end{align}
and 
 \begin{align}
         { \mathcal{F}^{---}(C) = \bigg(\frac{(123)^{2s_1}(1 21)^{2s_2-2s_1}(131)^{2s_3-2s_1} } {((\bar1 1\bar2)(23\bar3))^{s_2+s_3-s_1+\frac{1}{2}}}\bigg)} .
\end{align}
As one can see from the above results,  in going from \((++-)\) helicity to \((---)\) helicity,  one had to simply switch all \(2\) to \(\bar 2\) and \(\bar 2\) to \(2\) in the minors,  and switch all \(\xi_2\) to \(\bar\xi_2\). This procedure is general,  and one can obtain results in any other helicity starting from a given expression by simply switching the barred quantities to unbarred and vice versa.\\
\textit{Four-point functions—}\\
 At four points,  helicity flipping for \(\mathcal{F}(C)\) is non-trivial,  and one has to work out things carefully to get the results. Nevertheless,  the expression for \(\hat\delta(C\cdot \Xi)\) follows the helicity-flipping procedure like that of two- and three-point, and as an example,  the \(\delta(C\cdot \Xi^{++++})\) is given by 
\begin{align}\label{GDelta4ptInPPPP}
  \hat{\delta}(C\cdot\Xi^{++++})= &-\frac{1}{4}\big((\bar1\Bar{2}12)+(\bar3\Bar{4}34)+(\bar1\bar313)+(\Bar{2}\Bar{4}24)+(\bar1\Bar{4} 14)+(\Bar{2}\bar323)\big)\notag\\&-\frac{{\xi}_1\xi_2}{2}\big((\bar 1\Bar{2} \bar 33)+(\bar 1\Bar{2}\Bar{4}4)\big)-\frac{{\xi}_1{\xi}_3}{2}\big((\bar1\bar3\Bar{2}2)+(\bar1\bar3\Bar{4}4)\big)-\frac{{\xi}_1\xi_4}{2}\big((\bar1\Bar{4}\Bar{2}2)+(\bar1\Bar{4}\bar3{3})\big)\notag\\
  &-\frac{\xi_2{\xi}_3}{2}\big((\Bar{2}\bar3 \Bar{4}4)+(\Bar{2}\bar3\bar1 {1})\big)-\frac{\xi_2\xi_4}{2}\big((\Bar{2}\Bar{4}\bar3{3})+(\Bar{2}\Bar{4}\bar1{1})\big)-\frac{{\xi}_3\xi_4}{2}\big((\bar3\Bar{4}\bar1{1})+(\bar3\Bar{4}\Bar{2}2)\big)\notag\\
  &+{\xi}_1\xi_2{\xi}_3\xi_4(\bar1\Bar{2}\bar3\Bar{4}), 
\end{align}
and one can get the other helicity $\hat\delta(C\cdot\Xi^{h_1\cdots h_n})$ by using helicity flipping.


\bibliography{biblio}
\bibliographystyle{JHEP}
\end{document}